\documentclass[onecolumn]{aastex62}

\usepackage{epsfig}
\usepackage{epstopdf}
\usepackage{graphics}
\usepackage{amsmath}
\usepackage{xcolor}
\usepackage{aas_macros}

\usepackage{array}
\usepackage{booktabs}
\usepackage{amsmath,amssymb,amsfonts,amsbsy}
\usepackage{mathrsfs}
\usepackage{url}
\usepackage{multirow}
\usepackage{xcolor}
\usepackage{color}
\usepackage{ulem}
\usepackage{enumerate}

\renewcommand{\vec}[1]{\boldsymbol{#1}}

\newcolumntype{p}{D{,}{\pm}{-1}}

\renewcommand{\figurename}{Fig.}
\renewcommand{\tablename}{Tab.}

\renewcommand{\arraystretch}{1.5}

\begin{document}

\title{Large-scale anisotropy of Galactic cosmic rays as a probe of local cosmic-ray propagation}
% related to local regular magnetic field}

\correspondingauthor{Wei Liu, Yi-Qing Guo}
\email{liuwei@ihep.ac.cn, guoyq@ihep.ac.cn}

\author{Ai-feng Li}
 \affiliation{college of Information Science and Engineering, Shandong Agricultural University, Taian 271018, China}%Lines break automatically or can be forced with \\
\author{Qiang Yuan}
\affiliation{Key Laboratory of Dark Matter and Space Astronomy, Purple Mountain Observatory, Chinese Academy of Sciences, Nanjing 210008, China}
\affiliation{School of Astronomy and Space Science, University of Science and Technology of China, Hefei 230026, China}
\author{Wei Liu}
\affiliation{Key Laboratory of Particle Astrophysics, Institute of High Energy Physics, Chinese Academy of Sciences, Beijing 100049, China}
\author{Yi-qing Guo}
\affiliation{Key Laboratory of Particle Astrophysics, Institute of High Energy Physics, Chinese Academy of Sciences, Beijing 100049, China}
\affiliation{University of Chinese Academy of Sciences, Beijing 100049, China}

%\date{\today}% It is always \today, today,
             %  but any date may be explicitly specified
%We study the influence of anisotropic diffusion on the phase.
% to the LRMF
%If the perpendicular diffusion is still much slower, the phase above $100$ TeV still points toward the LRMF, but a revise direction applies as indicated by the KASCADE experiment. If the perpendicular diffusion grows faster than the parallel one, so that the diffusion is approximately isotropic above $\sim 100$ TeV, the phase turns to the Galactic center, which is consistent with the As$\gamma$ and ICECUBE observations.

\begin{abstract}
Recent studies have shown that the anisotropy is of great value to decipher cosmic rays' origin and propagation. We have built an unified scenario to describe the observations of the energy spectra and the large-scale anisotropy and called attention to their synchronously evolution with energy. In this work, the impact of of the local regular magnetic field (LRMF) and corresponding anisotropic diffusion on large-scale anisotropy have been investigated. When the perpendicular diffusion coefficient is much smaller than the parallel one, the dipole anisotropy points to the LRMF and the observational phase below $100$ TeV could be reproduced. Moreover we find that the dipole phase above $100$ TeV strongly depends on the evolution of local diffusion. But the current measurements at that energy are still scarce. We suggest that more precise measurements at that energy could be carried out to unveil the local diffusion and further the local turbulence.
%The current observational data of dipole anisotropy have large inconsistencies above $100$ TeV.
\end{abstract}
%investigate the influence of LRMF and corresponding local anisotropy diffusion on the dipole anisotropy.
%\pacs{Valid PACS appear here}% PACS, the Physics and Astronomy
                             % Classification Scheme.

%\keywords{cosmic rays supernova remnants}%Use showkeys class option if keyword display desired

%\maketitle

%%%%%%%%%%%%%%%%%%%%%%%%%%%%%%%%%%%%%%%%%%%%%%%%%%%%%%%%%%%%%%%%%%%%%%
\section{Introduction}
\label{sec:intro}
%%%%%%%%%%%%%%%%%%%%%%%%%%%%%%%%%%%%%%%%%%%%%%%%%%%%%%%%%%%%%%%%%%%%%%

% with energy
% revealed
%provided the high-precision and two dimensional observations of CR anisotropy. Tiny CR anisotropy with relative amplitudes is at the order of $10^{-4} \sim 10^{-3}$.
%	only energy spectrum information could not
%focus on
%of different compositions
%distinguish
% for a long time
%With more and more observations with high precision, p
%The past researches are chiefly concerned with the features of energy spectra. There is an increasing realization that the energy spectra alone could not rule out various theoretical models effectively. Recent
%The observations indicate that
%The amplitude initially grows up with energy gradually up to $\sim 10$ TeV and then decreases until $\sim 100$ TeV, later increases again above that energy.
% flip
%with high-precision

The understanding of the transport process of cosmic rays (CRs) in Galaxy is crucial to uncover their origin. The past researches are chiefly concerned with the features of energy spectra. Now more and more studies have shown that the CR anisotropy could be another useful observable. Due to the diffusive propagation, the arrival directions of CRs are approximately isotropic in observation. However the mass of evidence has identified that the distribution of the arrival direction is uneven, with the relative intensity varying from $\sim 10^{-4}$ to $\sim 10^{-2}$, which is called anisotropy. So far a lot of experiments have measured the energy dependence of the dipole anisotropy in a wide energy range  \citep{2006Sci...314..439A, 2010ApJ...711..119A, 2017ApJ...836..153A, 2007PhRvD..75f2003G, 2008PhRvL.101v1101A, 2009ApJ...698.2121A, 2010ApJ...718L.194A, 2011ApJ...740...16A, 2012ApJ...746...33A, 2013ApJ...765...55A, 2016ApJ...826..220A, 2013PhRvD..88h2001B, 2015ApJ...809...90B, 2014ApJ...796..108A, 2019ApJ...871...96A}. The results show that its evolution with energy is non-trivial. Less than $1$ PeV, the dipole amplitude grows up with energy below $10$ TeV and above $100$ TeV respectively, whereas between $10$ TeV and $100$ TeV, it indicates a downward trend. What's more, the phase points toward $\sim 3$ hrs below $100$ TeV, which evidently deviates from the Galactic center (which is at about $-6$ hrs). And above that energy, a flip occurs undoubtedly.

Actually, in the framework of the diffusion model, the uneven distribution of the overall CR sources predicts a large-scale anisotropy. Under the isotropic diffusion, the amplitude of the dipole anisotropy is expected to rise with energy, which soon far exceeds the available measurements, up to two orders of magnitude at $\sim 100$ TeV, according to the diffusion coefficient inferred from the boron-to-carbon ratio. And the direction of dipole anisotropy aligns with the Galactic center immutably \citep{2012JCAP...01..011B}, which goes against with the measurements at low energy. Introducing a local CR source at the direction of anti-Galactic center could partially ease the tension. Since nearby the solar system, the approximation of continuous distribution is no longer valid, the finite and discrete point-like CR sources have to be considered properly. \cite{2011JCAP...02..031M} and \cite{2012A&A...544A..92B} proposed that the finite number of CR sources around the solar system could give rise to the large fluctuations in the observed energy spectrum at high energy, so that the spectral hardening of CR nuclei above $200$ GV could originate from the local source. Furthermore when a local CR source is located at the direction of anti-Galactic center, the magnitude of dipole anisotropy could reduce to the observational level, through cancelling the streaming from the background CR sources \citep{2017PhRvD..96b3006L}. However, since the contribution from the local source drops off sharply at $\sim 100$ TeV, the magnitude raises again and outruns observations. Meanwhile the problem of the phase flip is still unsolved.
%can not be explained
%The CR anisotropy is attributed to the following reasons.
%Young nearby sources, such as SNRs, can influence the spatial gradient of CR particles in the solar system\citep{2006APh....25..183E, 2012JCAP...01..010B}, therefore anisotropy may result from nearby sources.
%, in which the diffusion coefficient nearby the Galactic plane is much slower
%smaller due to the larger turbulence induced by the local supernova explosions
%the amplitude of 

Another solution for this issue is to modified the diffusion coefficient, i.e. the so-called spatial-dependent propagation (SDP). It was initially proposed to explain the spectral hardening of CR nuclei above $200$ GV \citep{2012ApJ...752L..13T, 2016ApJ...819...54G, 2018ApJ...869..176L}. Compared with the conventional diffusion model, the diffusion process nearby the Galactic disk is supposed to be much slower. Thereupon the dipole anisotropy could reduce visibly to the observational level. But since the anisotropy is caused by the large-scale distribution of CR sources, the phase points to the Galactic center as well. In the recent works, we established an unified picture to describe energy spectra and anisotropy \citep{2019JCAP...10..010L, 2019JCAP...12..007Q} by introducing a local source in the SDP model. We proposed that the spectral features and dipole anisotropy have a common origin and they synchronously evolve with energy. The excess of nuclei between $200$ GeV and $\sim 20$ TeV and the anisotropy below $100$ TeV are dominated by a local source. Beyond 100 TeV, the background sources override in both energy spectra and dipole anisotropy. The phase at lower energy indicates that the position of local source is close to the direction of anti-Galatic center and far from the Galactic disk. We suggest that the Geminga SNR at its pulsar's birth place could be a plausible candidate \citep{2022ApJ...926...41Z}. Similar scenarios have also been offered by \citet{2021PhRvD.104j3013F} and \cite{2022MNRAS.511.6218Z} recently.

%the amplitude variation and phase flipping in the dipole anisotropy have a common origin with the spectral hardening of nuclei above $200$ GeV and ensuing falloff at $\sim 20$ TeV. Less than $100$ TeV, and the spectral features are . 
%features
%transition
%Then, we focus on calculating the amplitude and phase of anisotropy from GeV-PeV energy region. It is well known that the anisotropy of CRs is proportional to the spatial gradient of the CR density and the diffusion coefficient.
%We find the Geminga SNR is still the most likely candidate of local source.
%When the perpendicular and parallel diffusion are close, the phase points toward the Galactic center. When
%the perpendicular diffusion is much slower than the parallel one,
%We further study the energy dependence of the parallel and perpendicular diffusion coefficients according to the phase of CR anisotropy.
%But the phase less than $100$ TeV could not be well accounted for.
% local ordered magnetic field on distance scales less than 0.1 pc can be inferred from the emission of energetic neutral atoms (ENA) from the outer heliosphere observed by the Interstellar Boundary Explorer (IBEX)
%show that the phase less than $100$ TeV is consistent with the direction of the local regular magnetic field \cite{}.

In the past years, the study of the emission of energetic neutral atoms by the IBEX experiment revealed that there is a local regular magnetic field within $0.1$ pc around the solar system, with the direction at $l \sim 210.5^\circ$ and $b \simeq -57.1^\circ$. The direction of LRMF is coincident with the dipole phase below $100$ TeV. The studies of \cite{2014Sci...343..988S}, \cite{2016PhRvL.117o1103A} and \cite{2020ApJ...892....6L} demonstrated that this originates from the anisotropic diffusion which guides the CRs propagate along the LRMF. In this work, we focus on the influence of the LRMF and the corresponding anisotropic diffusion on the dipole anisotropy. At lower energy, when the perpendicular diffusion is much smaller, the CRs principally propagate along the LRMF. The dipole phase directs to the LRMF, which well conforms with the measurements. But above $100$ TeV, the phase is found to be dependent on the energy dependence of the perpendicular diffusion. If the perpendicular diffusion is still smaller than parallel, the diffusion of CRs is anisotropic and the dipole phase just flips $180^\circ$. But if the perpendicular diffusion grows faster and is comparable to the parallel above $\sim 100$ TeV, the diffusion turns to be isotropic, and the phase directs to the Galactic center. Thereupon, we demonstrate that the dipole phase could be applied to ascertain the local diffusion. However current observations beyond tens of TeV are still lack and the available data have large uncertainties. The precise measurements by e.g. LHAASO, HAWC and ICECUBE experiments, could help determine the evolution of local diffusion with energy and further enhance our understanding of the local turbulence.

%, so that both cases could not be clearly distinguished
%Both ratio of perpendicular-to-parallel diffusion coefficient $\varepsilon$ and difference of power indexes $\Delta \delta = \delta_\perp -\delta_\parallel$ are comprehensively investigated. With perpendicular diffusion decreasing, i.e. $\varepsilon$ gradually diminishes, the whole amplitude of anisotropy reduces. Meanwhile the dipole phase of anisotropy less than $100$ TeV changes from the local source to LRMF. When $\Delta \delta \approx 0$, namely the perpendicular diffusion is still much smaller than parallel diffusion above $100$ TeV, the phase of anisotropy at that energy range would flip over $180$ degrees, i.e. points toward the reverse direction of the local magnetic field. But if $\Delta \delta > 0$, i.e. the perpendicular diffusion grows faster than parallel one so that both values are very close above $100$ TeV, the phase would point toward to the Galactic center. 

%%%%%%%%%%%%%%%%%%%%%%%%%%%%%%%%%%%%%%%%%%%%%%%%%%%%%%%%%%%%%%%%%%%%%%
\section{Model Description}
\label{sec:model}
%%%%%%%%%%%%%%%%%%%%%%%%%%%%%%%%%%%%%%%%%%%%%%%%%%%%%%%%%%%%%%%%%%%%%%

%%%%%%%%%%%%%%%%%%%%%%%%%%%%%%%%%%%%%%%%%%%%%%%%%%%%%%%%%%%%%%%%%%%%%%
\subsection{Spatially-dependent propagation} \label{subsec:SDP}
%%%%%%%%%%%%%%%%%%%%%%%%%%%%%%%%%%%%%%%%%%%%%%%%%%%%%%%%%%%%%%%%%%%%%%
%about a few kpc
%	have recently that the diffusion coefficient of CRs near the galactic disk
%Especially, the local diffusion coefficient nearby the source unveiled by the HAWC experiment is nearly two orders of magnitude smaller than the global one inferred by the B/C ratio \citep{2017Sci...358..911A}. The SDP model
% \citep{2011Sci...332...69A}

In recent years, the spatial-dependent propagation (SDP) model has been given more and more attention. It was initially introduced as a Two Halo model \citep{2012ApJ...752L..13T} to explain the excess of primary proton and helium fluxes above $200$ GeV. Later, it was further used to account for the excess of secondary and heavier components \citep{2015PhRvD..92h1301T, 2016PhRvD..94l3007F, 2016ApJ...819...54G, 2018ApJ...869..176L, 2020ChPhC..44h5102T, 2020FrPhy..1624501Y}, diffuse gamma-ray distribution \citep{2018PhRvD..97f3008G} and large-scale anisotropy \citep{2019JCAP...10..010L, 2019JCAP...12..007Q, 2022ApJ...926...41Z}. The recent measurement of TeV halo around the pulsars found that the CRs diffuse significantly slower than the inference from the boron-to-carbon ratio, which strongly support the assumption that diffusion could be spatial-dependent \citep{2017Sci...358..911A, 2021PhRvL.126x1103A}.
%at particles diffuse significantly slower around pulsar
%For a comprehensive introduction to the SDP model, one can refer to \cite{2016ApJ...819...54G} and \cite{2018ApJ...869..176L}.

%spatial dependent and
%The Galactic disk and its surrounding area are called the inner halo (IH) region, where the diffusion is much slower and relevant to the radial distribution of CR sources.
% The extensive diffusive region outside the IH is named as the outer halo (OH) region, in which the diffusion is only rigidity dependent. 

In the SDP model, the whole diffusive halo is divided into two regions. Nearby the Galactic disk and its surrounding area are called the inner halo (IH) region, the turbulence level is affected by the activities of supernova explosions immensely. Hence it is expected to be intense near the large population of sources and correspondingly the diffusion process is slower, while at regions with fewer sources, the turbulence is moderate and the diffusion tends to be fast. Thus, the diffusion coefficient in the inner halo is relevant to the radial distribution of CR sources. Far away from the Galactic disk, i.e. so-called outer halo (OH) region, the turbulence is less impacted by the stellar activities and believed to be self-generated by CRs themselves, so the diffusion approaches to be only rigidity-dependent.

In this work, the half thickness of the whole diffusive halo is defined as $z_h$, and IH and OH regions are $\xi z_{h}$ and $(1-\xi) z_{h}$ respectively. Both $z_h$ and $\xi$ are determined by fitting B/C ratio and nuclei spectra. The diffusion coefficient $D_{xx}$ is parameterized as:
\begin{equation}
D_{xx}(r,z, {\cal R} )= D_{0}F(r,z)\beta^{\eta} \left(\dfrac{\cal R}{{\cal R}_{0}} \right)^{\delta_0 F(r,z)} ~,
\label{eq:diffusion}
\end{equation}
in which the reference rigidity ${\cal R}_{0}$ is fixed to $4$ GV. The spatial dependence $F(r,z)$ is written as
\begin{equation}
F(r,z) =
\begin{cases}
g(r,z) +\left[1-g(r,z) \right] \left(\dfrac{z}{\xi z_0} \right)^{n} , &  |z| \leqslant \xi z_h \\
1 ~, & |z| > \xi z_h
\end{cases}.
\end{equation}
$g(r,z) = N_m/[1+f(r,z)]$, with $f(r,z)$ representing the spatial distribution of CR sources. The CR sources are approximated as axisymmetric-distributed, i.e. $f(r,z) \propto (r/r_\odot)^{\alpha} \exp[-\beta (r -r_\odot)/r_\odot] \exp(-|z|/z_s)$, where $r_\odot = 8.5$ kpc and $z_s = 0.2$ kpc. $\alpha$ and $\beta$ are taken as $1.69$ and $3.33$ \citep{1996A&AS..120C.437C}.

The injection spectrum of background sources is assumed to have a form of power-law of rigidity with an exponential cutoff, namely
\begin{equation}
Q({\cal R}) \propto {\cal R}^{-\nu} \exp \left(-\dfrac{{\cal R}}{{\cal R}_{\rm c} } \right) ~.
\end{equation}
The diffusion-reacceleration (DR) model is adopted in the propagation equation and the numerical package, DRAGON, is used to compute the background CR distribution \citep{1475-7516-2008-10-018}.

%The region where CRs diffuse in the Milky Way is called a magnetic halo, which is usually approximated as a cylinder with its radial boundary equal to the Galactic radius, i.e. $R = 20$ kpc. The half thickness $z_h$ is usually determined by fitting the B/C ratio along with diffusion coefficient \citep {2007ARNPS..57..285S}. Both CR sources and the interstellar medium are usually assumed to be concentrated near the Galactic disk, whose average thickness $z_s$  is roughly $200$ pc.

%Therefore, we work in aSDP frame \citep {2012ApJ...752L..13T, 2016ApJ...819...54G, 2016ChPhC..40a5101J}, whose diffusion coefficient are different in inner halo and outer halo. The parameterized diffusion coefficient we adopt is  \citep {2018PhRvD..97f3008G, 2018ApJ...869..176L}.

\subsection{Local source}
%In general, the distribution of cosmic ray sources are not smooth, in particular in our local environment. The presence of local and young CR sources can introduce modulations of phase and amplitude, even if their individual contribution to total CR flux is only subdominant.

In our previous works \citep{2019JCAP...10..010L, 2019JCAP...12..007Q, 2022ApJ...926...41Z}, the spectral anomaly above $200$ GV and dipole anisotropy below $100$ TeV are attributed to Geminga SNR, which is the remnant after the explosion of Geminga's progenitor. Its characteristic age is inferred from the spin-down luminosity of Geminga pulsar, which is about $\tau = 3.4 \times 10^5$ years \citep{2005AJ....129.1993M}. The position is backtracked to $l = 194.3^\circ, b = -13^\circ$, and the distance to the solar system is $\sim 330$ pc \citep{2007Ap&SS.308..225F}. The CR flux from the Geminga SNR can be computed by solving the time-dependent diffusion equation using the Green's function method, assuming an boundary at infinity \citep{2017PhRvD..96b3006L, 2019JCAP...10..010L}. As for the instantaneous and point-like injection, the spatial distribution is 
\begin{equation}
\psi(r,{\cal R},t)=\frac{q_{\rm inj}({\cal R})}{(\sqrt{2\pi}\sigma)^3}
\exp\left(-\frac{(\vec{r}-\vec{r}^\prime)^2}{2\sigma^2}\right),
\end{equation}
where $\sigma({\cal R},t)=\sqrt{2D_{xx}({\cal R})t}$ is the effective diffusion length within time $t$. The injection spectrum $q_{\rm inj}({\cal R})$ is parameterized as a power-law function of rigidity with an exponential cutoff, i.e. $q_0{\cal R}^{-\alpha} \exp(-{\cal R}/{\cal R}'_{\rm c})$.

%The diffusion coefficient $D({\cal R})$ is adopted as the value nearby the solar system.
%The normalization $q_0$ at $1$ GV is obtained by fitting the CR energy spectra.
%where $q_{\rm inj}({\cal R})$ is the injection spectrum of local source.

%amplitude of the
%with observed relative difference between the fluxes in the maximum and
%minimum directions $\phi_{\rm max}$ and $\phi_{\rm max}$
%as \citep{2015PhRvL.114b1101M}
% \dfrac{(\phi_{\rm max}-\phi_{\rm min})}{(\phi_{\rm max}+\phi_{\rm min})}
%its vector form can be
%%diffusion of CRs consists of  two components: parallel  and  perpendicular to the magnetic field.

%with an uncertainty of $\sim 1.5^\circ$
%on distance scales
%by observing the emission of energetic neutral atoms (ENA) from the outer heliosphere
%, with 
% and could be the origin of low-energy phase
% within $40$ pc
%when the LRMF is taken into account, 

\subsection{Anisotropic diffusion}
The IBEX experiment has detected a regular magnetic field with spatial scale $\sim 0.1$ pc around the solar system \citep{2009Sci...326..959M, 2013ApJ...776...30F}. The magnetic field axis is along the direction of $l \simeq 210.5^\circ, b \simeq -57.1^\circ$. Similar finding was reported from the polarization measurements of the local stars \citep{2015ApJ...814..112F}. This direction is coincident with the dipole phase below $100$ TeV \citep{2016PhRvL.117o1103A}. Since that the spatial scale of LRMF is much smaller than the average propagation length inferred from the boron-to-carbon ratio, the LRMF does not have remarkable impact on the energy spectra. But CRs diffuse anisotropically in the local interstellar space under the influence of LRMF and the arrival direction and the corresponding dipole anisotropy is expected to be modified correspondingly. In this case, the diffusion coefficient $D$ is replaced by the tensor $D_{ij}$, which is written as
\begin{equation}
D_{ij}\,\equiv\,D_\perp\delta_{ij}\,+\,\big(D_\|-D_\perp\big)b_ib_j ~, ~ ~
b_i = \dfrac{B_i}{|\vec{B}|}
\label{eq:D_ij_1}
\end{equation}
where $b_i$ is the $i$-th component of the unit vector of LRMF \citep{1999ApJ...520..204G, 2017JCAP...10..019C}. $D_{\parallel}$ and $D_{\perp}$ are the diffusion coefficients parallel and perpendicular to the LRMF, respectively. In this work, each of them is parameterized as a power-law function of rigidity,
\begin{align}
D_\parallel &= D_{0\parallel} \beta^{\eta} \left(\frac{\cal R}{{\cal R}_0} \right)^{\delta_\|} ~, \\
D_\perp &=\,D_{0\perp} \beta^{\eta} \left(\frac{\cal R}{{\cal R}_0} \right)^{\delta_\perp} = \varepsilon D_{0\parallel} \beta^{\eta} \left(\frac{\cal R}{{\cal R}_0} \right)^{\delta_\perp} ~.
\label{eq:DparaDperp}
\end{align}
Here $\varepsilon = \dfrac{D_{0\perp}}{D_{0\parallel} }$ is the ratio between perpendicular and parallel diffusion coefficient at the reference rigidity ${\cal R}_{0}$. When taking the local anisotropic diffusion into account, the dipole anisotropy becomes accordingly
\begin{equation}
\vec{\delta} = \dfrac{3\vec{D} \cdot \nabla \psi}{v \psi} = \dfrac{3}{v \psi} D_{ij}  \dfrac{ \partial \psi}{ \partial x_j} ~.
\end{equation}

\section{Results and Discussion}
\label{sec:results}

First of all, the propagation parameters can be determined by fitting the boron-to-carbon ratio. The LRMF has the spatial scale of $\sim 0.1$ pc \citep{2013ApJ...776...30F}, which is much less than the average propagation length of CRs. The propagated energy spectra is hardly impacted by the LRMF. Hence the CR fluxes from both background and local sources could be evaluated in terms of SDP. In the SDP model, the unknown propagation parameters are $D_0$, $\delta_0$, $N_m$, $\xi$, $n$, $v_A$ and $z_h$, whose values are shown in table \ref{tab:transport}. In addition, the normalization, power index and cut-off rigidity of background (local) nuclei $i$ in the injection spectra, i.e. $A^{\rm i}$, $\nu^{\rm i}$ and ${\cal R}_c$ ($q^{\rm i}_{0}, \alpha^{\rm i}$ and ${\cal R}^\prime_c$) have to be settled by fitting the energy spectra. The cutoff rigidities of different compositions are regarded as the limits of acceleration in the sources and assumed to be $Z$-dependent.

\begin{figure*}[!ht]
	\includegraphics[width=0.48\textwidth]{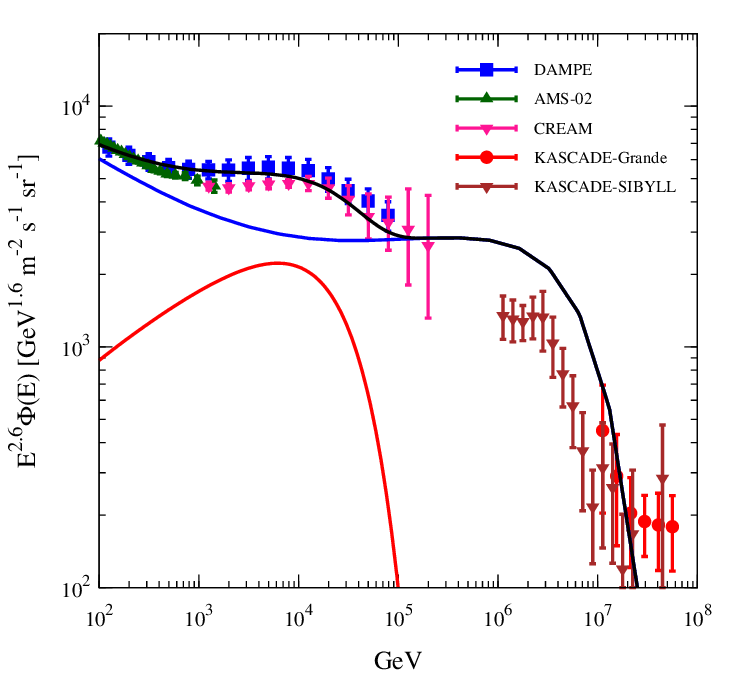}
	\includegraphics[width=0.48\textwidth]{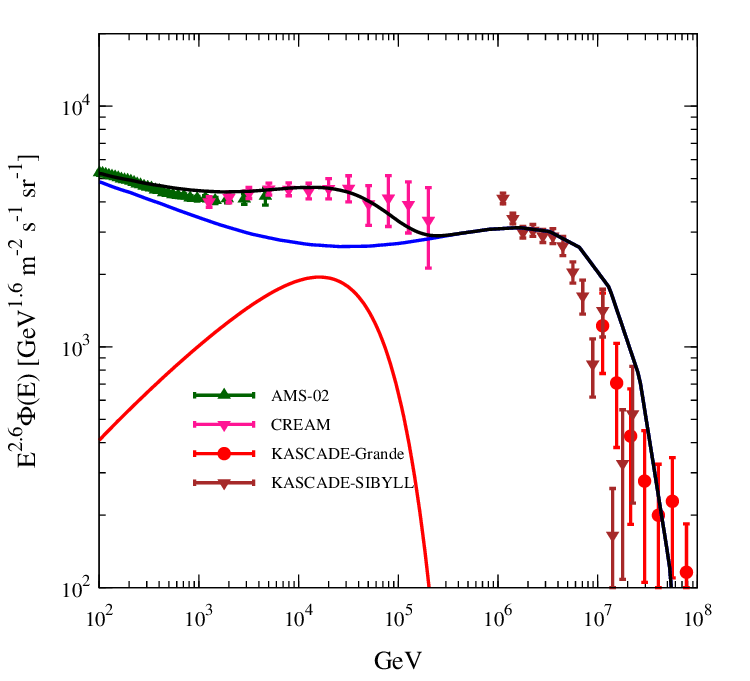}
	\caption{Energy spectra of protons (left) and helium nuclei (right). The data
		points are taken from
		DAMPE\citep{2019SciA....5.3793A, 2021PhRvL.126t1102A},
		AMS-02 \citep{2015PhRvL.114q1103A, 2017PhRvL.119y1101A},
		CREAM-III \citep{2017ApJ...839....5Y}, NUCLEON \citep{2017JCAP...07..020A}, KASCADE \citep{2005APh....24....1A} and KASCADE-Grande \citep{2013APh....47...54A} respectively.  The blue lines are the background fluxes, and the red lines are the fluxes from a nearby Geminga SNR source respectively. The black lines represent the total fluxes.
	}
	\label{fig:Fig1}
\end{figure*}

\begin{figure*}
	\centering
	\includegraphics[width=0.48\textwidth]{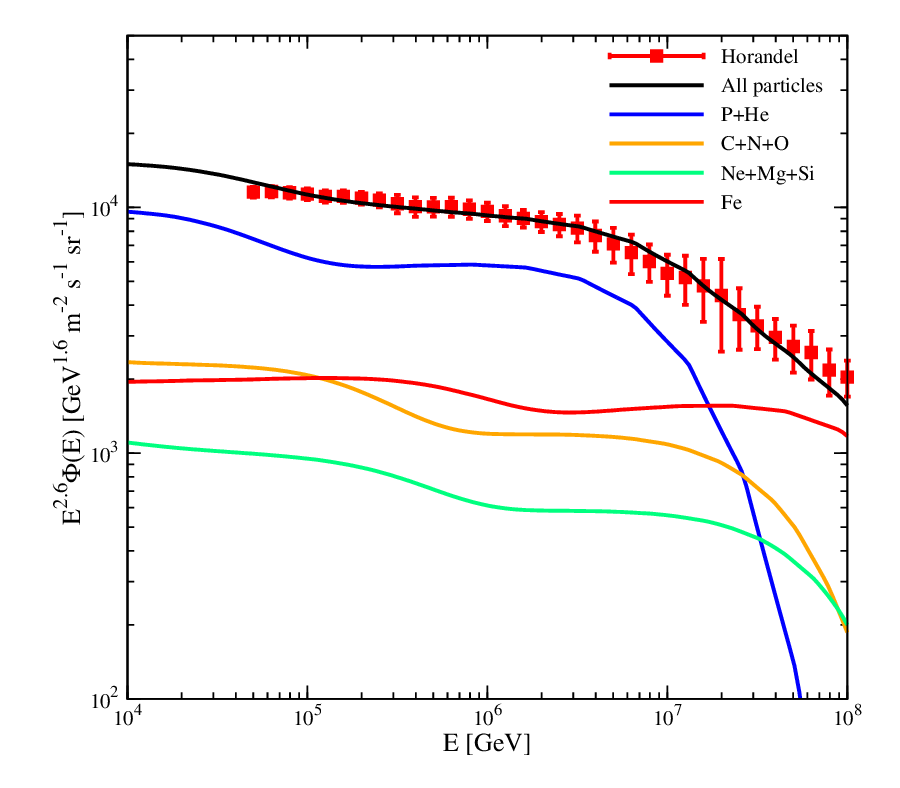}
	\caption{The all-particle spectra multiplied by $E^{2.6}$.
		The data points are taken from \citep{2003APh....19..193H}.
		The solid lines with different colors are the model predictions of different mass groups, and the black solid line is the total contribution.
	}
	\label{fig:all_spec}
\end{figure*}

Fig. \ref{fig:Fig1} shows the fitting to the proton (left), helium (right) spectra, in which red, blue and black lines are the contribution from local source, background sources and sum of all respectively. And Fig. \ref{fig:all_spec} illustrates the fitting to the all-particle spectra. The injection parameters of major compositions are listed in table \ref{tab:para_inj}. To account for the softening at tens of TeV in proton and helium spectra, the cut-off rigidity of the local source is $25$ TV. And to describe the all-particle spectrum as well as cut-offs of proton and helium at PeV energies, the cut-off rigidity of the background sources are $7$ PV.

%Fig. \ref{fig:ratio} shows the comparison of the B/C and ${}^{10}$Be/${}^9$Be ratios between the SDP predictions and the data. The values of corresponding propagation parameters are listed in table \ref{tab:transport}. The red lines is the B/C ratio computed only from background sources, and the black line is the one with additional carbon contribution from local source. As can be seen that, the carbon flux from the local source lowers the total B/C ratio above $\sim 10$ GeV. Within the uncertainty of the measurements, the computed B/C ratio is still consistent with the latest AMS-02 measurement. Due to lack of the precise observation, the measurements of ${}^{10}$Be/${}^9$Be have large errors, and our fitting could also account for the current data.

\begin{table}
	\begin{center}
		\caption{Fitted SDP parameters.}
		\begin{tabular}{|ccccccccc|}
			\hline
			& $D_0$~~~    &   $\delta_0$~~~     &    $N_{\rm m}$~~~    &    $\xi$~~~   &    $n$~~~  &    $v_A$~~~    &    $z_h$ &  \\
			& [${\rm cm}^2 \cdot {\rm s}^{-1}$] & & & & & [${\rm km}\cdot{\rm s}^{-1}$] & [kpc]  &\\
			\hline
			& $4.87 \times 10^{28}$  & 0.58  & 0.62    & 0.1   & 4          & 6        &  5    &      \\
			\hline
		\end{tabular}
		\label{tab:transport}
	\end{center}
\end{table}

\begin{table*}
	\begin{center}
		\begin{tabular}{|c|ccc|ccc|}
			\hline
			& \multicolumn{3}{c|}{Background} & \multicolumn{3}{c|}{Local source} \\
			\hline
			%\toprule[1.5pt]
			Element & A$^\dagger$ & ~~~$\nu$~~~  & ~~~$\mathcal R_{c}$~~~ & ~~~$q_0$~~~~~ & ~~~~~$\alpha$~~~ & ~~~${\cal R}'_c$~~~ \\
			\hline
			& $[({\rm m}^2\cdot {\rm sr}\cdot {\rm s}\cdot {\rm GeV})^{-1}]$ & & [PV] & [GeV$^{-1}$] & &  [TV] \\
			\hline
			p   & $1.91\times 10^{-2}$    & 2.34   &  7  & $8.28\times 10^{52}$  & 2.16 & 25 \\
			He & $1.43\times 10^{-3}$   & 2.27     &  7  & $2.35\times 10^{52}$  & 2.08  &  25  \\
			C   & $6.15\times 10^{-5}$   & 2.31    &  7  & $7.2\times 10^{50}$    & 2.13 &  25  \\
			N   & $7.67\times 10^{-6}$   & 2.34    &  7  & $1.13\times 10^{50}$  & 2.13 &   25  \\
			O   & $8.20\times 10^{-5}$   & 2.36    &  7  & $1.11\times 10^{51}$ & 2.13  &   25  \\
			Ne & $8.05\times 10^{-6}$   & 2.28   &  7  & $1.13\times 10^{50}$ & 2.13  &  25 \\
			Mg & $1.62\times 10^{-5}$   & 2.39     &  7  & $1.08\times 10^{50}$ & 2.13  &  25  \\
			Si & $1.28\times 10^{-5}$     & 2.37   &  7  & $1.05\times 10^{50}$ & 2.13  &   25  \\
			Fe & $1.23\times 10^{-5}$    & 2.29    &  7  & $2.20\times 10^{50}$ & 2.13   &  25  \\
			\hline
			%\bottomrule[1.5pt]
		\end{tabular}\\
		$^\dagger${The normalization is set at total energy $E = 100$ GeV.}
	\end{center}
	\caption{Fitted injection parameters of the background and local sources.}
	\label{tab:para_inj}
\end{table*}

%First, we calculate the proton and helium spectra. Figure \ref{fig:Fig1} shows the energy spectra of protons and helium obtained for the model calculation, with the red, blue and black lines representing the contributions from the nearby source, the background sources and the sum of them, respectively. It can be seen that the contribution of nearby Geminga SNR source can simultaneously account for the spectral hardening features at $\sim$ 200 GeV, and softening features at  $\sim$ 10 TeV.
%Through adding different compositions together, we  get the all-particle spectrum as shown in  Figure \ref{fig:all_spec}, which is well consistent with the observational data.

%The anisotropy of CRs depends on the sum of the CR flows from the background  and the local source. %after being anisotropically diffused in the LRMF.
%When calculating anisotropy,
%Since there are four parameters in the parallel and perpendicular diffusion coefficients, which could not
%In order tosimplify the calculation, let's first set $\Delta\delta=\delta_\perp-\delta_\parallel=0$, where $\Delta\delta$ is the difference between  $\delta_\perp$ and $\delta_\parallel$.
%Despite that the measurements still have uncertainties, we assume t
%\citep{2014Sci...343..988S, 2016PhRvL.117o1103A}
% and the corresponding parameters are those in Table \ref{tab:transport}
%, while the perpendicular diffusion is to be investigated

%  is defined to be along $l = 210.5^\circ$ and $b = -57.1^\circ$ \citep{2009Sci...326..959M}.
%scale down the perpendicular diffusion by 

\begin{figure*}
\includegraphics[width=0.98\textwidth]{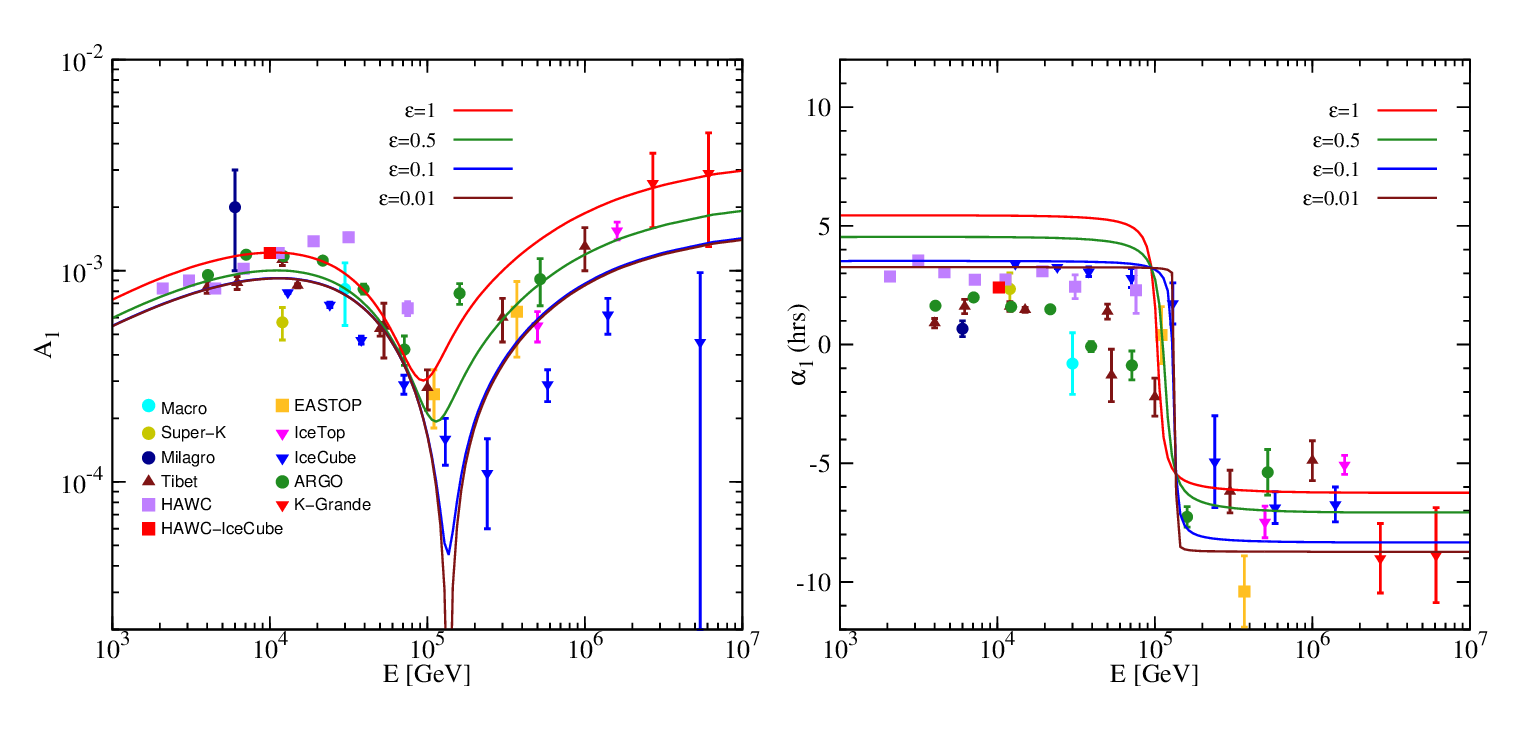}
\caption{The energy dependence of the amplitude (left) and phase (right) of the dipole anisotropy when $\varepsilon=1, ~0.5, ~0.1, ~0.01$ respectively with $\delta_\parallel =\delta_\perp$.
The data points are taken from Marco \citep{2003PhRvD..67d2002A}, SuperKamiokande \citep{2007PhRvD..75f2003G}, EAS-TOP \citep{1995ICRC....2..800A, 1996ApJ...470..501A, 2009ApJ...692L.130A},
Milagro  \citep{2009ApJ...698.2121A}, IceCube  \citep{2010ApJ...718L.194A, 2012ApJ...746...33A},
Ice-Top \citep{2013ApJ...765...55A}, ARGO-YBJ \citep{2015ApJ...809...90B}, Tibet \citep{2005ApJ...626L..29A, 2010ApJ...711..119A, 2017ApJ...836..153A}, KASCADE-Grande \citep{2015ICRC...34..281C}, HAWC \citep{2019ApJ...871...96A}, HAWC-IceCube \citep{2019ApJ...871...96A}
}
\label{fig:amp_ph}
\end{figure*}

Unlike the energy spectra, the LRMF could dramatically affect the arrival direction of CRs and thus the dipole anisotropy by altering the direction of CR streaming. To study the influence of the LRMF, the anisotropic diffusion in the local environment and thus the anisotropy are investigated. The value measured by the IBEX is chosen as the direction of LRMF. And the parallel diffusion is equal to the diffusion in the SDP. Fig. \ref{fig:amp_ph} shows the dipole amplitude and phase by varying $\varepsilon$ from $1$ to $0.01$ with $\delta_\perp = \delta_\parallel$. With $\varepsilon$ decreasing, the perpendicular diffusion thus diminishes gradually, which engenders the descent of the overall dipole amplitude. Meanwhile the smaller $\varepsilon$ is, the deeper the trench at $\sim 100$ TeV becomes. This can be understood as follows. The dipole amplitude depends on the CR flow $\vec{D} \cdot \nabla \psi$. When $\varepsilon \ll 1$, only the CR flow projected along the LRMF play a leading role, while the CR flow perpendicular to the LRMF is suppressed due to the decrease of the perpendicular diffusion. And due to the lack of the CR flow perpendicular to the LRMF, the trench at $\sim 100$ TeV becomes smaller, namely sharper.

%This can be understood as follows. The dipole amplitude depends on the CR flow $\vec{D} \cdot \nabla \psi$. When $\varepsilon = 1$, i.e. the diffusion is isotropic, the direction of CR flow depends on the density gradient. The background flow propagates along the Galactic disk and points to the Galactic center. \textcolor{red}{At the same time, the local source is at  $l=194.3^\circ, b=-13^\circ$, which is beneath the Galactic disk. The CR flow from the local source is thus decomposed into two parts: one is along the Galactic disk and the other is perpendicular to the Galactic disk. And in doing so, the net flow consists of two parts: the flow along the Galactic disk and the vertical one.} \textcolor{red}{the diffusion becomes anisotropic, both background and local CR flows are apt to propagate along the LRMF, but with opposite direction. Thus the part projected onto the LRMF dominates.} In this case, the net flow $\vec{D} \cdot \nabla \psi$ and the dipole magnitude diminishes. 

More important, $\varepsilon$ also revises the phase, as shown in right of Fig. \ref{fig:amp_ph}. Less than 100 TeV, the local source, i.e. Geminga SNR, dominates. When $\varepsilon$ closes to $1$, the diffusion is approximately isotropic and the dipole phase points to the position of Geminga SNR. Above 100 TeV, the background sources dominates and the phase turns to the Galactic center. This corresponds to the regular isotropic diffusion, which is shown as the red line. With the decrease of $\varepsilon$, the dipole phase gradually changes. When $\varepsilon$ is less than $0.1$, the CR flow tends to propagate along LRMF more, and the phase directs to the LRMF less than $100$ TeV, which is compatible with the measurements. At higher energy, the background streaming is dominated. The phase makes a $180^\circ$ flip and points the opposite direction of LRMF, as shown by the blue and brown solid lines. It is worth noting that the transition of dipole phase at $\sim 100$ TeV also becomes sharper with $\varepsilon$ decreasing due to lack of the component perpendicular to the LRMF. 

%Since both background and local CR flows are along LRMF, the orientation has only two directions. Therefore a break occurs at $\sim 100$ TeV. Compared with smaller $\varepsilon$, the transition is modest when $\varepsilon$ closes to $1$. However the measurements are still inconsistent at that energy. The precise measurements could help determine the local diffusion.

%and the accessible amplitude of anisotropy at $100$ TeV is much smaller than the case of $\varepsilon = 1$.
% and along the LRMF
%comes from the direction of
%, whereas t
% meets the minimum

%It's worth noting that the trench at $100$ TeV becomes deeper.

% shows the comparison of the anisotropy of CRs with different $\varepsilon$.
%To study the induced anisotropy changes of CRs by the ratio of  perpendicular and parallel diffusion coefficients,
%the anisotropy of different $\varepsilon$ (1, 0.5,  0.1,  0.01) at   $\Delta\delta = 0$ are calculated and are shown in Figure \ref{fig:delta0}.
%Figure \ref{fig:diffepsilon} shows the anisotropies, where different types of dotted lines represent different different  $\varepsilon$.
%If  $\varepsilon=1$, $D_\perp=D_\parallel$, CRs are isotropic diffusion.
%Thus, the phase points toward the nearby Geminga SNR source below 100 TeV, while points to the GC above 100 TeV.
%As $\varepsilon$ decreases from 1 to 0, the phase gradually transitions from the nearby source  to the LRMF below 100 TeV, meanwhile  shifts from GC to anti LRMF above 100 TeV.
%As can be seen from Figure \ref{fig:diffepsilon}, if $\varepsilon$

\begin{figure*}
	\includegraphics[width=0.98\textwidth]{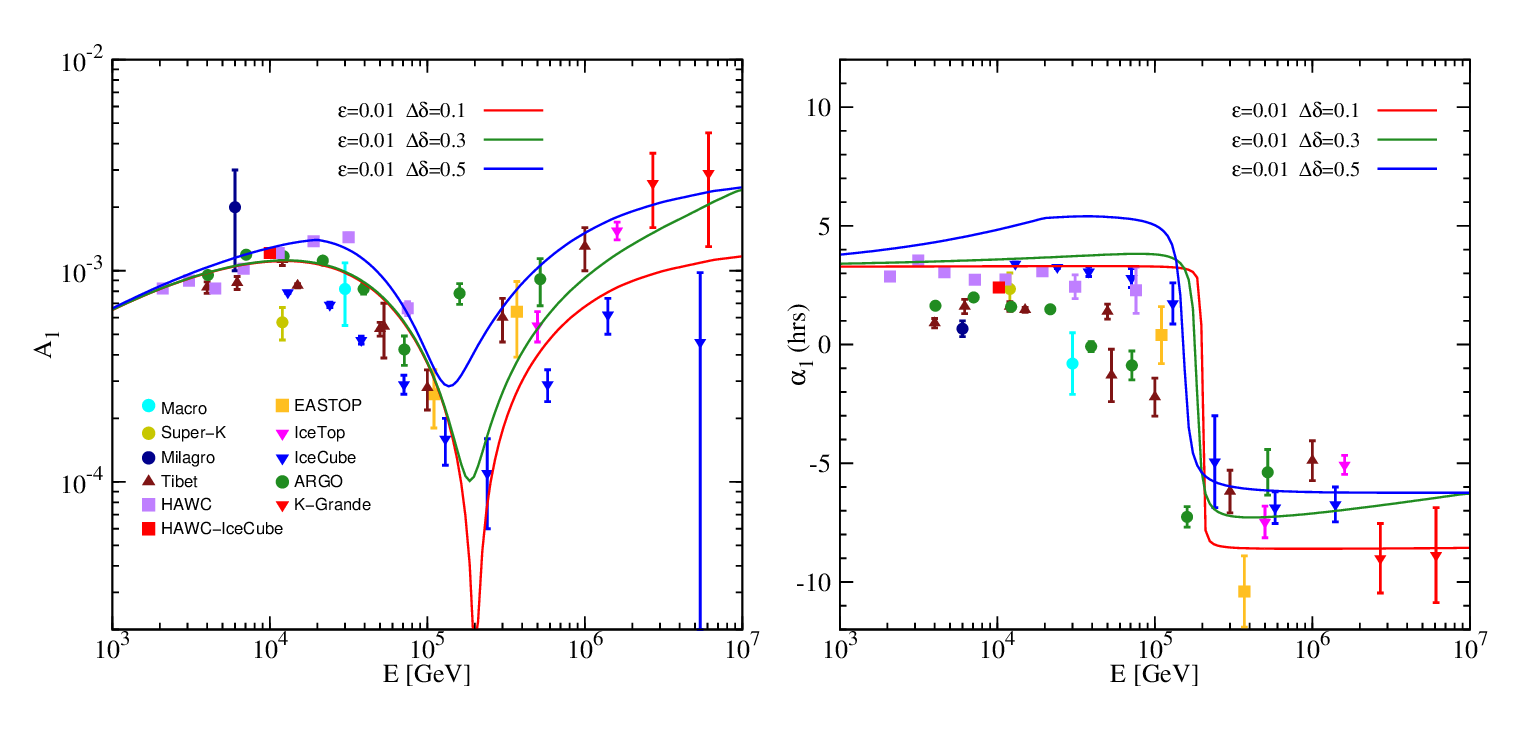}
	\caption{Energy dependences of the amplitude (left) and phase (right) of the anisotropies for $\varepsilon = 0.01$. The three black lines correspond to $\Delta\delta = \delta_\perp-\delta_\parallel = 0.1,~0.3, ~0.5$ respectively.
	}
	\label{fig:epsilon_001}
\end{figure*}

Fig. \ref{fig:epsilon_001} illustrates the influence of $\delta_\perp$ on dipole anisotropy by defining $\Delta\delta = \delta_\perp - \delta_\parallel$. Here $\epsilon$ is fixed to $0.01$ so that the dipole phase orients the LRMF at lower energy. But at higher energy, the dipole anisotropy depends on $\Delta\delta$. When $\Delta\delta$ is close $0$, for example $0.1$, the perpendicular diffusion grows slower with energy, so that above hundreds of TeV it is still smaller than the parallel one, the phase points to the opposite direction of LRMF, as the red solid shows. But when $\Delta\delta = 0.3$, the perpendicular diffusion grows faster than the parallel and both of them becomes comparable above 100 TeV. In other words, the diffusion converts to be isotropic at that energy. Hence above $100$ TeV, the amplitude increases and the orientation changes from LRMF to the Galactic center, see the green line. There is a transition of direction above 100 TeV, from the IRMF to the Galactic center. When the perpendicular diffusion grows much faster than the parallel, for example $\Delta\delta = 0.5$, the diffusion approaches to be isotropic before $100$ TeV and the phase changes from the LRMF to the local source. This is disfavored by the available measurements and thus excluded. We conclude that the dipole anisotropy could be applied to constrain the evolution of diffusion process with energy.

% And above 100 TeV the phase points to the Galactic center.

\begin{figure*}
	\includegraphics[width=0.48\textwidth]{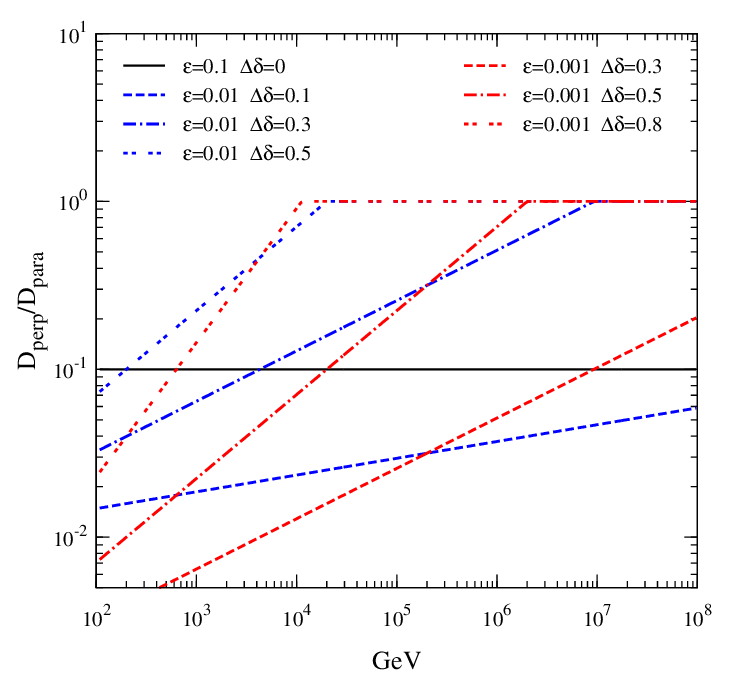}
	\caption{ Energy dependences of ${D_\perp}/{D_\parallel}$. The black line is the result for $\varepsilon = 0.01$ and $\Delta\delta =0$. The three blue broken lines correspond to $\Delta\delta =0.1,~0.3, ~0.5$ with $\varepsilon = 0.01$. The three red broken lines correspond to $\Delta\delta =0.3,~0.5, ~0.8$ with $\varepsilon = 0.001$.
	}
	\label{fig:Diffusion}
\end{figure*}

In Figure \ref{fig:Diffusion}, we illustrate the energy dependence of the ratio ${D_\perp}/{D_\parallel}$ by varying $\Delta \delta$ with $\varepsilon = 0.01$ (blue) and $\varepsilon = 0.001$ (red) respectively. The black solid line corresponds to $\varepsilon = 0.1$ and $\delta_\perp = \delta_\parallel$, i.e. the blue solid line in Figure \ref{fig:amp_ph}. When the ${D_\perp}$ is always smaller than ${D_\parallel}$ , for example the blue ($\Delta \delta = 0.1, \varepsilon = 0.01$) and red ($\Delta \delta = 0.3, \varepsilon = 0.001$) dash lines, the dipole phase is always along LRMF, but with a $180^\circ$ reversal above $100$ TeV. When $\Delta \delta = 0.3, \varepsilon = 0.01$ and $\Delta \delta = 0.5, \varepsilon = 0.001$, i.e. blue and red dash-dot lines, both perpendicular and parallel diffusion are of comparable above $100$ TeV, the phase changes gradually from LRMF to the Galactic center. When $\Delta \delta$ is too large, for example $0.5$ when $\varepsilon = 0.01$ and $0.8$ when $\varepsilon = 0.001$, the diffusion approaches the isotropic below $100$ TeV, the phase points from the LRMF to the local source less than 100 TeV, and turn to the Galactic center at higher energy. Thus the precise measurements of the dipole anisotropy constrain the available range of $\delta_\perp$ when fixing $\varepsilon$. The smaller $\varepsilon$ is, the larger the available range of $\delta_\perp$ becomes.

However the observations of the anisotropy above $100$ TeV are still scarce. The measurements by the AS$\gamma$ and ICECUBE experiments showed that the phase points to the Galactic center \citep{2016ApJ...826..220A, 2017ApJ...836..153A, 2018ApJ...861...93B}. The KASCADE-Grande's measurement indicated an excess in the anti-LRMF direction at the corresponding energy range \citep{2015ICRC...34..281C, 2016PhRvL.117o1103A}. However, the latest analysis using more KASCADE-Grande data still did not report the dipole anisotropy with high enough significance \citep{2019ApJ...870...91A, 2019ApJ...886L..18A}. Nevertheless, that the dipole phase above $100$ TeV directs to the Galactic center is still not well established, subject to the small data sample and worse energy resolution. Meanwhile due to the limited energy resolution of ground-base instruments at high energy, the transition of dipole phase above 100 TeV, as the green solid line in Fig. \ref{fig:epsilon_001} shows, is hard to measure. We hope the precise measurements of the anisotropy with higher energy resolution, for example LHAASO, HAWC and ICECUBE experiments, could untangle that confusion and impose constraints on the energy dependence of parallel and perpendicular diffusion.
%is not prominent and only gives an upper limit

\section{Summary}
\label{sec:summary}

The CR anisotropies have been observed for a long time. Recently there is an increasing realization that the large-scale dipole anisotropy could unveil the nearby sources. In the previous works, we built up a self-consistent propagation scenario to explain both the spectral hardening in the energy spectra and the evolution of the dipole anisotropy. In this work, the local regular magnetic field and the corresponding anisotropic diffusion are introduced to study their impact  on the dipole anisotropy. The ratio of perpendicular-to-parallel diffusion coefficient at reference rigidity and difference of their power indexes are investigated respectively. 

%The magnitude of perpendicular diffusion affects the amplitude of anisotropy.

As the perpendicular diffusion diminishes, the CR fluxes are apt to propagate along the LRMF nearby the solar system. Therefore the dipole phase changes from the position of the local source to the direction of the LRMF less than $100$ TeV, which is more consistent with the current observations. And above $100$ TeV, the phase makes a $180^\circ$ turnaround and points to the direction of the LRMF. And we notice that the amplitude of anisotropy could reduce due to the decrease of perpendicular diffusion. %, which seems to conform with the observation of KASCADE-Grande

However AS$\gamma$, ARGO, HAWC and ICECUBE experiments indicate that above $100$ TeV, the dipole phase points to the Galactic center. Under the scenario of LRMF, it inevitably hints a transition of diffusion mechanism, in which  diffusion turns from anisotropic to isotropic with energy increasing. That means the perpendicular diffusion grows faster than parallel diffusion as energy increases. We study the power index of the perpendicular diffusion compared to the parallel one. The dipole phase between tens of TeV and several PeV could effectively constrain the power index of the perpendicular diffusion for the certain ratio of perpendicular-to-parallel diffusion coefficient at reference rigidity. The measurements of CR anisotropy, for example LHAASO, HAWC and ICECUBE experiments, are expected to help determine the diffusion coefficients and could constrain the local turbulence.

Moreover, we want to call attention that due to the poor knowledge of the detector response, it is hard to directly compare the  anisotropy intensities at the various declination bands. That is, the ground-based experiments are incapable of observing CR anisotropies along the Earth's rotation axis. The available observations of anisotropy is still incomplete. The observations just indicate the right ascension below $100$ TeV is coincident with the LRMF. To confirm that, a real 2D observation containing declination information is still necessary. To do this, the  detector response should be simulated at least to the level $\sim 10^{-3}$. We also hope the space instruments could perform the accurate observations in the near future.

Recently \cite{2017ApJ...835..258G} investigated the influence of the turbulence on the large-scale anisotropy. They found that the specific turbulence in the local interstellar environment could distort dipole anisotropy by introducing higher order component and reshape the distribution of right ascension. Meanwhile \cite{2022ApJ...927..110K} studied the impact of the specific turbulence on formation of the small-scale anisotropy. In the future work, we would further study how the specific turbulence affects the diffusion and anisotropy.

\section*{Acknowledgements}
This work is supported by the National Key R$\&$D Program of China grant (2018YFA0404202) and  the National Natural Science Foundation of China (11963004, 11635011, 11875264, U1831208, U1738205, U2031110) and Shandong Province Natural Science Foundation ( ZR2020MA095).

%%%%%%%%%%%%%%%%%%%%%%%%%%%%%%%%%%%%%%%%%%%%%%%%%%%%%%%%%%%%%%%%%%%%%%
\bibliographystyle{apj}
\bibliography{ref1}

\begin{thebibliography}{}
\expandafter\ifx\csname natexlab\endcsname\relax\def\natexlab#1{#1}\fi

\bibitem[{{Aartsen} {et~al.}(2013){Aartsen}, {Abbasi}, {Abdou}, {Ackermann},
  {Adams}, {Aguilar}, {Ahlers}, {Altmann}, {Andeen}, {Auffenberg}, \&
  et~al.}]{2013ApJ...765...55A}
{Aartsen}, M.~G., {Abbasi}, R., {Abdou}, Y., {et~al.} 2013, apj, 765, 55

\bibitem[{{Aartsen} {et~al.}(2016){Aartsen}, {Abraham}, {Ackermann}, {Adams},
  {Aguilar}, {Ahlers}, {Ahrens}, {Altmann}, {Anderson}, {Ansseau}, \&
  et~al.}]{2016ApJ...826..220A}
{Aartsen}, M.~G., {Abraham}, K., {Ackermann}, M., {et~al.} 2016, apj, 826, 220

\bibitem[{{Abbasi} {et~al.}(2010){Abbasi}, {Abdou}, {Abu-Zayyad}, {Adams},
  {Aguilar}, {Ahlers}, {Andeen}, {Auffenberg}, {Bai}, {Baker}, \&
  et~al.}]{2010ApJ...718L.194A}
{Abbasi}, R., {Abdou}, Y., {Abu-Zayyad}, T., {et~al.} 2010, apj, 718, L194

\bibitem[{{Abbasi} {et~al.}(2011){Abbasi}, {Abdou}, {Abu-Zayyad}, {Adams},
  {Aguilar}, {Ahlers}, {Altmann}, {Andeen}, {Auffenberg}, {Bai}, \&
  et~al.}]{2011ApJ...740...16A}
---. 2011, apj, 740, 16

\bibitem[{{Abbasi} {et~al.}(2012){Abbasi}, {Abdou}, {Abu-Zayyad}, {Ackermann},
  {Adams}, {Aguilar}, {Ahlers}, {Allen}, {Altmann}, {Andeen}, \&
  et~al.}]{2012ApJ...746...33A}
---. 2012, apj, 746, 33

\bibitem[{{Abdo} {et~al.}(2008){Abdo}, {Allen}, {Aune}, {Berley}, {Blaufuss},
  {Casanova}, {Chen}, {Dingus}, {Ellsworth}, {Fleysher}, {Fleysher},
  {Gonzalez}, {Goodman}, {Hoffman}, {H{\"u}ntemeyer}, {Kolterman}, {Lansdell},
  {Linnemann}, {McEnery}, {Mincer}, {Nemethy}, {Noyes}, {Pretz}, {Ryan},
  {Parkinson}, {Shoup}, {Sinnis}, {Smith}, {Sullivan}, {Vasileiou}, {Walker},
  {Williams}, \& {Yodh}}]{2008PhRvL.101v1101A}
{Abdo}, A.~A., {Allen}, B., {Aune}, T., {et~al.} 2008, Physical Review Letters,
  101, 221101

\bibitem[{{Abdo} {et~al.}(2009){Abdo}, {Allen}, {Aune}, {Berley}, {Casanova},
  {Chen}, {Dingus}, {Ellsworth}, {Fleysher}, {Fleysher}, {Gonzalez}, {Goodman},
  {Hoffman}, {Hopper}, {H{\"u}ntemeyer}, {Kolterman}, {Lansdell}, {Linnemann},
  {McEnery}, {Mincer}, {Nemethy}, {Noyes}, {Pretz}, {Ryan}, {Parkinson},
  {Shoup}, {Sinnis}, {Smith}, {Sullivan}, {Vasileiou}, {Walker}, {Williams}, \&
  {Yodh}}]{2009ApJ...698.2121A}
{Abdo}, A.~A., {Allen}, B.~T., {Aune}, T., {et~al.} 2009, apj, 698, 2121

\bibitem[{{Abeysekara} {et~al.}(2014){Abeysekara}, {Alfaro}, {Alvarez},
  {{\'A}lvarez}, {Arceo}, {Arteaga-Vel{\'a}zquez}, {Ayala Solares}, {Barber},
  {Baughman}, {Bautista-Elivar}, {Belmont}, {BenZvi}, {Berley}, {Bonilla
  Rosales}, {Braun}, {Caballero-Mora}, {Carrami{\~n}ana}, {Castillo}, {Cotti},
  {Cotzomi}, {de la Fuente}, {De Le{\'o}n}, {DeYoung}, {Diaz Hernandez},
  {D{\'{\i}}az-V{\'e}lez}, {Dingus}, {DuVernois}, {Ellsworth}, {Fiorino},
  {Fraija}, {Galindo}, {Garfias}, {Gonz{\'a}lez}, {Goodman}, {Gussert},
  {Hampel-Arias}, {Harding}, {H{\"u}ntemeyer}, {Hui}, {Imran}, {Iriarte},
  {Karn}, {Kieda}, {Kunde}, {Lara}, {Lauer}, {Lee}, {Lennarz}, {Le{\'o}n
  Vargas}, {Linnemann}, {Longo}, {Luna-Garc{\'{\i}}a}, {Malone}, {Marinelli},
  {Marinelli}, {Martinez}, {Martinez}, {Mart{\'{\i}}nez-Castro}, {Matthews},
  {McEnery}, {Mendoza Torres}, {Miranda-Romagnoli}, {Moreno}, {Mostaf{\'a}},
  {Nellen}, {Newbold}, {Noriega-Papaqui}, {Oceguera-Becerra}, {Patricelli},
  {Pelayo}, {P{\'e}rez-P{\'e}rez}, {Pretz}, {Rivi{\`e}re}, {Rosa-Gonz{\'a}lez},
  {Ruiz-Velasco}, {Ryan}, {Salazar}, {Salesa Greus}, {Sandoval}, {Schneider},
  {Sinnis}, {Smith}, {Sparks Woodle}, {Springer}, {Taboada}, {Toale},
  {Tollefson}, {Torres}, {Ukwatta}, {Villase{\~n}or}, {Weisgarber},
  {Westerhoff}, {Wisher}, {Wood}, {Yodh}, {Younk}, {Zaborov}, {Zepeda}, {Zhou},
  \& {HAWC Collaboration}}]{2014ApJ...796..108A}
{Abeysekara}, A.~U., {Alfaro}, R., {Alvarez}, C., {et~al.} 2014, apj, 796, 108

\bibitem[{{Abeysekara} {et~al.}(2017){Abeysekara}, {Albert}, {Alfaro},
  {Alvarez}, {{\'A}lvarez}, {Arceo}, {Arteaga-Vel{\'a}zquez}, {Avila Rojas},
  {Ayala Solares}, {Barber}, {Bautista-Elivar}, {Becerril}, {Belmont-Moreno},
  {BenZvi}, {Berley}, {Bernal}, {Braun}, {Brisbois}, {Caballero-Mora},
  {Capistr{\'a}n}, {Carrami{\~n}ana}, {Casanova}, {Castillo}, {Cotti},
  {Cotzomi}, {Couti{\~n}o de Le{\'o}n}, {De Le{\'o}n}, {De la Fuente},
  {Dingus}, {DuVernois}, {D{\'\i}az-V{\'e}lez}, {Ellsworth}, {Engel},
  {Enr{\'\i}quez-Rivera}, {Fiorino}, {Fraija}, {Garc{\'\i}a-Gonz{\'a}lez},
  {Garfias}, {Gerhardt}, {Gonz{\'a}lez Mu{\~n}oz}, {Gonz{\'a}lez}, {Goodman},
  {Hampel-Arias}, {Harding}, {Hern{\'a}ndez}, {Hern{\'a}ndez-Almada}, {Hinton},
  {Hona}, {Hui}, {H{\"u}ntemeyer}, {Iriarte}, {Jardin-Blicq}, {Joshi},
  {Kaufmann}, {Kieda}, {Lara}, {Lauer}, {Lee}, {Lennarz}, {Vargas},
  {Linnemann}, {Longinotti}, {Luis Raya}, {Luna-Garc{\'\i}a}, {L{\'o}pez-Coto},
  {Malone}, {Marinelli}, {Martinez}, {Martinez-Castellanos},
  {Mart{\'\i}nez-Castro}, {Mart{\'\i}nez-Huerta}, {Matthews}, {Mirand
  a-Romagnoli}, {Moreno}, {Mostaf{\'a}}, {Nellen}, {Newbold}, {Nisa},
  {Noriega-Papaqui}, {Pelayo}, {Pretz}, {P{\'e}rez-P{\'e}rez}, {Ren}, {Rho},
  {Rivi{\`e}re}, {Rosa-Gonz{\'a}lez}, {Rosenberg}, {Ruiz-Velasco}, {Salazar},
  {Salesa Greus}, {Sand oval}, {Schneider}, {Schoorlemmer}, {Sinnis}, {Smith},
  {Springer}, {Surajbali}, {Taboada}, {Tibolla}, {Tollefson}, {Torres},
  {Ukwatta}, {Vianello}, {Weisgarber}, {Westerhoff}, {Wisher}, {Wood},
  {Yapici}, {Yodh}, {Younk}, {Zepeda}, {Zhou}, {Guo}, {Hahn}, {Li}, \&
  {Zhang}}]{2017Sci...358..911A}
{Abeysekara}, A.~U., {Albert}, A., {Alfaro}, R., {et~al.} 2017, Science, 358,
  911

\bibitem[{{Abeysekara} {et~al.}(2019){Abeysekara}, {Alfaro}, {Alvarez},
  {Arceo}, {Arteaga-Vel{\'a}zquez}, {Avila Rojas}, {Belmont-Moreno}, {BenZvi},
  {Brisbois}, {Capistr{\'a}n}, {Carramiana}, {Casanova}, {Cotti}, {Cotzomi},
  {D{\'\i}az-V{\'e}lez}, {De Le{\'o}n}, {De la Fuente}, {Dichiara},
  {DuVernois}, {Espinoza}, {Fiorino}, {Fleischhack}, {Fraija},
  {Galv{\'a}n-G{\'a}mez}, {Garc{\'\i}a-Gonz{\'a}lez}, {Gonz{\'a}lez},
  {Goodman}, {Hampel-Arias}, {Harding}, {Hernandez}, {Hona},
  {Hueyotl-Zahuantitla}, {Iriarte}, {Jardin-Blicq}, {Joshi}, {Lara}, {Le{\'o}n
  Vargas}, {Luis-Raya}, {Malone}, {Marinelli}, {Mart{\'\i}nez-Castro},
  {Martinez}, {Matthews}, {Miranda-Romagnoli}, {Moreno}, {Mostaf{\'a}},
  {Nellen}, {Newbold}, {Nisa}, {Noriega-Papaqui}, {P{\'e}rez-P{\'e}rez},
  {Pretz}, {Ren}, {Rho}, {Rivi{\`e}re}, {Rosa-Gonz{\'a}lez}, {Rosenberg},
  {Salazar}, {Salesa Greus}, {Sandoval}, {Schneider}, {Schoorlemmer}, {Sinnis},
  {Smith}, {Surajbali}, {Taboada}, {Tollefson}, {Torres}, {Villaseor},
  {Weisgarber}, {Wood}, {Zepeda}, {Zhou}, {{\'A}lvarez}, {HAWC Collaboration},
  {Aartsen}, {Ackermann}, {Adams}, {Aguilar}, {Ahlers}, {Ahrens}, {Altmann},
  {Andeen}, {Anderson}, {Ansseau}, {Anton}, {Arg{\"u}elles}, {Auffenberg},
  {Axani}, {Backes}, {Bagherpour}, {Bai}, {Barbano}, {Barron}, {Barwick},
  {Baum}, {Bay}, {Beatty}, {Becker Tjus}, {Becker}, {BenZvi}, {Berley},
  {Bernardini}, {Besson}, {Binder}, {Bindig}, {Blaufuss}, {Blot}, {Bohm},
  {B{\"o}rner}, {Bos}, {B{\"o}ser}, {Botner}, {Bourbeau}, {Bourbeau},
  {Bradascio}, {Braun}, {Bretz}, {Bron}, {Brostean-Kaiser}, {Burgman}, {Busse},
  {Carver}, {Cheung}, {Chirkin}, {Clark}, {Classen}, {Collin}, {Conrad},
  {Coppin}, {Correa}, {Cowen}, {Cross}, {Dave}, {Day}, {de Andr{\'e}}, {De
  Clercq}, {DeLaunay}, {Dembinski}, {Deoskar}, {De Ridder}, {Desiati}, {de
  Vries}, {de Wasseige}, {de With}, {DeYoung}, {D{\'\i}az-V{\'e}lez},
  {Dujmovic}, {Dunkman}, {Dvorak}, {Eberhardt}, {Ehrhardt}, {Eichmann},
  {Eller}, {Evenson}, {Fahey}, {Fazely}, {Felde}, {Filimonov}, {Finley},
  {Franckowiak}, {Friedman}, {Fritz}, {Gaisser}, {Gallagher}, {Ganster},
  {Garrappa}, {Gerhardt}, {Ghorbani}, {Giang}, {Glauch}, {Gl{\"u}senkamp},
  {Goldschmidt}, {Gonzalez}, {Grant}, {Griffith}, {Haack}, {Hallgren}, {Halve},
  {Halzen}, {Hanson}, {Hebecker}, {Heereman}, {Helbing}, {Hellauer},
  {Hickford}, {Hignight}, {Hill}, {Hoffman}, {Hoffmann}, {Hoinka},
  {Hokanson-Fasig}, {Hoshina}, {Huang}, {Huber}, {Hultqvist}, {H{\"u}nnefeld},
  {Hussain}, {In}, {Iovine}, {Ishihara}, {Jacobi}, {Japaridze}, {Jeong},
  {Jero}, {Jones}, {Kalaczynski}, {Kang}, {Kappes}, {Kappesser}, {Karg},
  {Karle}, {Katz}, {Kauer}, {Keivani}, {Kelley}, {Kheirandish}, {Kim},
  {Kintscher}, {Kiryluk}, {Kittler}, {Klein}, {Koirala}, {Kolanoski},
  {K{\"o}pke}, {Kopper}, {Kopper}, {Koskinen}, {Kowalski}, {Krings}, {Kroll},
  {Kr{\"u}ckl}, {Kunwar}, {Kurahashi}, {Kyriacou}, {Labare}, {Lanfranchi},
  {Larson}, {Lauber}, {Leonard}, {Leuermann}, {Liu}, {Lohfink}, {Lozano
  Mariscal}, {Lu}, {L{\"u}nemann}, {Luszczak}, {Madsen}, {Maggi}, {Mahn},
  {Makino}, {Mancina}, {Mari{\c{s}}}, {Maruyama}, {Mase}, {Maunu}, {Meagher},
  {Medici}, {Meier}, {Menne}, {Merino}, {Meures}, {Miarecki}, {Micallef},
  {Moment{\'e}}, {Montaruli}, {Moore}, {Moulai}, {Nagai}, {Nahnhauer},
  {Nakarmi}, {Naumann}, {Neer}, {Niederhausen}, {Nowicki}, {Nygren}, {Obertacke
  Pollmann}, {Olivas}, {O'Murchadha}, {O'Sullivan}, {Palczewski}, {Pandya},
  {Pankova}, {Peiffer}, {Pepper}, {P{\'e}rez de los Heros}, {Pieloth}, {Pinat},
  {Pizzuto}, {Plum}, {Price}, {Przybylski}, {Raab}, {Rameez}, {Rauch},
  {Rawlins}, {Rea}, {Reimann}, {Relethford}, {Renzi}, {Resconi}, {Rhode},
  {Richman}, {Robertson}, {Rongen}, {Rott}, {Ruhe}, {Ryckbosch}, {Rysewyk},
  {Safa}, {Sanchez Herrera}, {Sandrock}, {Sandroos}, {Santander}, {Sarkar},
  {Sarkar}, {Satalecka}, {Schaufel}, {Schlunder}, {Schmidt}, {Schneider},
  {Schneider}, {Sch{\"o}neberg}, {Schumacher}, {Sclafani}, {Seckel},
  {Seunarine}, {Soedingrekso}, {Soldin}, {Song}, {Spiczak}, {Spiering},
  {Stachurska}, {Stamatikos}, {Stanev}, {Stasik}, {Stein}, {Stettner},
  {Steuer}, {Stezelberger}, {Stokstad}, {St{\"o}{\ss}l}, {Strotjohann},
  {Stuttard}, {Sullivan}, {Sutherland}, {Taboada}, {Tenholt}, {Ter-Antonyan},
  {Terliuk}, {Tilav}, {Toale}, {Tobin}, {T{\"o}nnis}, {Toscano}, {Tosi},
  {Tselengidou}, {Tung}, {Turcati}, {Turcotte}, {Turley}, {Ty}, {Unger},
  {Unland Elorrieta}, {Usner}, {Vandenbroucke}, {Van Driessche}, {van Eijk},
  {van Eijndhoven}, {Vanheule}, {van Santen}, {Vraeghe}, {Walck}, {Wallace},
  {Wallraff}, {Wandler}, {Wandkowsky}, {Watson}, {Weaver}, {Weiss}, {Wendt},
  {Werthebach}, {Westerhoff}, {Whelan}, {Whitehorn}, {Wiebe}, {Wiebusch},
  {Wille}, {Williams}, {Wills}, {Wolf}, {Wood}, {Wood}, {Woolsey}, {Woschnagg},
  {Wrede}, {Xu}, {Xu}, {Xu}, {Yanez}, {Yodh}, {Yoshida}, {Yuan}, \& {IceCube
  Collaboration}}]{2019ApJ...871...96A}
{Abeysekara}, A.~U., {Alfaro}, R., {Alvarez}, C., {et~al.} 2019, \apj, 871, 96

\bibitem[{{Aglietta} {et~al.}(1995){Aglietta}, {Alessandro}, {Antonioli},
  {Arneodo}, {Bergamasco}, {Bertaina}, {Bosio}, {Castellina}, {Castagnoli},
  {Chaivasa}, {Cini}, {D' Ettorre Piazzoli}, {Di Sciascio}, {Fulgione},
  {Galeotti}, {Ghia}, {Iacovacci}, {Mannocchi}, {Melagrana}, {Mengotti Silva},
  {Morello}, {Navarra}, {Riccati}, {Saavedra}, {Trinchero}, {Vallania}, \&
  {Vernetto}}]{1995ICRC....2..800A}
{Aglietta}, M., {Alessandro}, B., {Antonioli}, P., {et~al.} 1995, in
  International Cosmic Ray Conference, Vol.~2, International Cosmic Ray
  Conference, 800

\bibitem[{{Aglietta} {et~al.}(1996){Aglietta}, {Alessandro}, {Antonioli},
  {Arneodo}, {Bergamasco}, {Bertaina}, {Bosio}, {Castellina}, {Castagnoli},
  {Chiavassa}, {Cini Castagnoli}, {D'Ettorre Piazzoli}, {di Sciascio},
  {Fulgione}, {Galeotti}, {Ghia}, {Iacovacci}, {Mannocchi}, {Melagrana},
  {Mengotti Silva}, {Morello}, {Navarra}, {Riccati}, {Saavedra}, {Trinchero},
  {Vallania}, {Vernetto}, \& {EAS-Top Collaboration}}]{1996ApJ...470..501A}
{Aglietta}, M., {Alessandro}, B., {Antonioli}, P., {et~al.} 1996, \apj, 470,
  501

\bibitem[{{Aglietta} {et~al.}(2009){Aglietta}, {Alekseenko}, {Alessandro},
  {Antonioli}, {Arneodo}, {Bergamasco}, {Bertaina}, {Bonino}, {Castellina},
  {Chiavassa}, {D'Ettorre Piazzoli}, {Di Sciascio}, {Fulgione}, {Galeotti},
  {Ghia}, {Iacovacci}, {Mannocchi}, {Morello}, {Navarra}, {Saavedra},
  {Stamerra}, {Trinchero}, {Valchierotti}, {Vallania}, {Vernetto}, \&
  {Vigorito}}]{2009ApJ...692L.130A}
{Aglietta}, M., {Alekseenko}, V.~V., {Alessandro}, B., {et~al.} 2009, \apjl,
  692, L130

\bibitem[{{Aguilar} {et~al.}(2015){Aguilar}, {Aisa}, {Alpat}, {Alvino},
  {Ambrosi}, {Andeen}, {Arruda}, {Attig}, {Azzarello}, {Bachlechner}, {Barao},
  {Barrau}, {Barrin}, {Bartoloni}, {Basara}, {Battarbee}, {Battiston}, {Bazo},
  {Becker}, {Behlmann}, {Beischer}, {Berdugo}, {Bertucci}, {Bigongiari},
  {Bindi}, {Bizzaglia}, {Bizzarri}, {Boella}, {de Boer}, {Bollweg},
  {Bonnivard}, {Borgia}, {Borsini}, {Boschini}, {Bourquin}, {Burger}, {Cadoux},
  {Cai}, {Capell}, {Caroff}, {Casaus}, {Cascioli}, {Castellini}, {Cernuda},
  {Cerreta}, {Cervelli}, {Chae}, {Chang}, {Chen}, {Chen}, {Cheng}, {Chen},
  {Cheng}, {Chou}, {Choumilov}, {Choutko}, {Chung}, {Clark}, {Clavero},
  {Coignet}, {Consolandi}, {Contin}, {Corti}, {Gil}, {Coste}, {Creus},
  {Crispoltoni}, {Cui}, {Dai}, {Delgado}, {Della Torre}, {Demirk{\"o}z},
  {Derome}, {Di Falco}, {Di Masso}, {Dimiccoli}, {D{\'\i}az}, {von Doetinchem},
  {Donnini}, {Du}, {Duranti}, {D'Urso}, {Eline}, {Eppling}, {Eronen}, {Fan},
  {Farnesini}, {Feng}, {Fiandrini}, {Fiasson}, {Finch}, {Fisher},
  {Galaktionov}, {Gallucci}, {Garc{\'\i}a}, {Garc{\'\i}a-L{\'o}pez},
  {Gargiulo}, {Gast}, {Gebauer}, {Gervasi}, {Ghelfi}, {Gillard}, {Giovacchini},
  {Goglov}, {Gong}, {Goy}, {Grabski}, {Grandi}, {Graziani}, {Guandalini},
  {Guerri}, {Guo}, {Haas}, {Habiby}, {Haino}, {Han}, {He}, {Heil}, {Hoffman},
  {Hsieh}, {Huang}, {Huh}, {Incagli}, {Ionica}, {Jang}, {Jinchi}, {Kanishev},
  {Kim}, {Kim}, {Kirn}, {Kossakowski}, {Kounina}, {Kounine}, {Koutsenko},
  {Krafczyk}, {La Vacca}, {Laudi}, {Laurenti}, {Lazzizzera}, {Lebedev}, {Lee},
  {Lee}, {Leluc}, {Levi}, {Li}, {Li}, {Li}, {Li}, {Li}, {Li}, {Li}, {Li}, {Li},
  {Lim}, {Lin}, {Lipari}, {Lippert}, {Liu}, {Liu}, {Lolli}, {Lomtadze}, {Lu},
  {Lu}, {Lu}, {Luebelsmeyer}, {Luo}, {Lv}, {Majka}, {Ma{\~n}{\'a}},
  {Mar{\'\i}n}, {Martin}, {Mart{\'\i}nez}, {Masi}, {Maurin}, {Menchaca-Rocha},
  {Meng}, {Mo}, {Morescalchi}, {Mott}, {M{\"u}ller}, {Ni}, {Nikonov},
  {Nozzoli}, {Nunes}, {Obermeier}, {Oliva}, {Orcinha}, {Palmonari},
  {Palomares}, {Paniccia}, {Papi}, {Pauluzzi}, {Pedreschi}, {Pensotti},
  {Pereira}, {Picot-Clemente}, {Pilo}, {Piluso}, {Pizzolotto}, {Plyaskin},
  {Pohl}, {Poireau}, {Postaci}, {Putze}, {Quadrani}, {Qi}, {Qin}, {Qu},
  {R{\"a}ih{\"a}}, {Rancoita}, {Rapin}, {Ricol}, {Rodr{\'\i}guez},
  {Rosier-Lees}, {Rozhkov}, {Rozza}, {Sagdeev}, {Sandweiss}, {Saouter},
  {Sbarra}, {Schael}, {Schmidt}, {von Dratzig}, {Schwering}, {Scolieri}, {Seo},
  {Shan}, {Shan}, {Shi}, {Shi}, {Shi}, {Siedenburg}, {Son}, {Spada},
  {Spinella}, {Sun}, {Sun}, {Tacconi}, {Tang}, {Tang}, {Tang}, {Tao},
  {Tescaro}, {Ting}, {Ting}, {Tomassetti}, {Torsti}, {T{\"u}rko{\v{g}}lu},
  {Urban}, {Vagelli}, {Valente}, {Vannini}, {Valtonen}, {Vaurynovich},
  {Vecchi}, {Velasco}, {Vialle}, {Vitale}, {Vitillo}, {Wang}, {Wang}, {Wang},
  {Wang}, {Wang}, {Wang}, {Weng}, {Whitman}, {Wienkenh{\"o}ver}, {Wu}, {Wu},
  {Xia}, {Xie}, {Xie}, {Xiong}, {Xin}, {Xu}, {Xu}, {Yan}, {Yang}, {Yang}, {Ye},
  {Yi}, {Yu}, {Yu}, {Zeissler}, {Zhang}, {Zhang}, {Zhang}, {Zhang}, {Zheng},
  {Zhuang}, {Zhukov}, {Zichichi}, {Zimmermann}, {Zuccon}, {Zurbach}, \& {AMS
  Collaboration}}]{2015PhRvL.114q1103A}
{Aguilar}, M., {Aisa}, D., {Alpat}, B., {et~al.} 2015, \prl, 114, 171103

\bibitem[{{Aguilar} {et~al.}(2017){Aguilar}, {Ali Cavasonza}, {Alpat},
  {Ambrosi}, {Arruda}, {Attig}, {Aupetit}, {Azzarello}, {Bachlechner}, {Barao},
  {Barrau}, {Barrin}, {Bartoloni}, {Basara}, {Ba{\c{s}}e{\v{g}}mez-du Pree},
  {Battarbee}, {Battiston}, {Becker}, {Behlmann}, {Beischer}, {Berdugo},
  {Bertucci}, {Bindel}, {Bindi}, {de Boer}, {Bollweg}, {Bonnivard}, {Borgia},
  {Boschini}, {Bourquin}, {Bueno}, {Burger}, {Burger}, {Cadoux}, {Cai},
  {Capell}, {Caroff}, {Casaus}, {Castellini}, {Cervelli}, {Chae}, {Chang},
  {Chen}, {Chen}, {Chen}, {Cheng}, {Chou}, {Choumilov}, {Choutko}, {Chung},
  {Clark}, {Clavero}, {Coignet}, {Consolandi}, {Contin}, {Corti}, {Creus},
  {Crispoltoni}, {Cui}, {Dadzie}, {Dai}, {Datta}, {Delgado}, {Della Torre},
  {Demakov}, {Demirk{\"o}z}, {Derome}, {Di Falco}, {Dimiccoli}, {D{\'\i}az},
  {von Doetinchem}, {Dong}, {Donnini}, {Duranti}, {D'Urso}, {Egorov}, {Eline},
  {Eronen}, {Feng}, {Fiandrini}, {Fisher}, {Formato}, {Galaktionov},
  {Gallucci}, {Garc{\'\i}a-L{\'o}pez}, {Gargiulo}, {Gast}, {Gebauer},
  {Gervasi}, {Ghelfi}, {Giovacchini}, {G{\'o}mez-Coral}, {Gong}, {Goy},
  {Grabski}, {Grandi}, {Graziani}, {Guo}, {Haino}, {Han}, {He}, {Heil},
  {Hoffman}, {Hsieh}, {Huang}, {Huang}, {Huh}, {Incagli}, {Ionica}, {Jang},
  {Jia}, {Jinchi}, {Kang}, {Kanishev}, {Khiali}, {Kim}, {Kim}, {Kirn}, {Konak},
  {Kounina}, {Kounine}, {Koutsenko}, {Kulemzin}, {La Vacca}, {Laudi},
  {Laurenti}, {Lazzizzera}, {Lebedev}, {Lee}, {Lee}, {Leluc}, {Li}, {Li}, {Li},
  {Li}, {Li}, {Li}, {Li}, {Lim}, {Lin}, {Lipari}, {Lippert}, {Liu}, {Liu},
  {Lordello}, {Lu}, {Lu}, {Luebelsmeyer}, {Luo}, {Luo}, {Lyu}, {Machate},
  {Ma{\~n}{\'a}}, {Mar{\'\i}n}, {Martin}, {Mart{\'\i}nez}, {Masi}, {Maurin},
  {Menchaca-Rocha}, {Meng}, {Mikuni}, {Mo}, {Mott}, {Nelson}, {Ni}, {Nikonov},
  {Nozzoli}, {Oliva}, {Orcinha}, {Palmonari}, {Palomares}, {Paniccia},
  {Pauluzzi}, {Pensotti}, {Perrina}, {Phan}, {Picot-Clemente}, {Pilo},
  {Pizzolotto}, {Plyaskin}, {Pohl}, {Poireau}, {Quadrani}, {Qi}, {Qin}, {Qu},
  {R{\"a}ih{\"a}}, {Rancoita}, {Rapin}, {Ricol}, {Rosier-Lees}, {Rozhkov},
  {Rozza}, {Sagdeev}, {Schael}, {Schmidt}, {Schulz von Dratzig}, {Schwering},
  {Seo}, {Shan}, {Shi}, {Siedenburg}, {Son}, {Song}, {Tacconi}, {Tang}, {Tang},
  {Tescaro}, {Ting}, {Ting}, {Tomassetti}, {Torsti}, {T{\"u}rko{\v{g}}lu},
  {Urban}, {Vagelli}, {Valente}, {Valtonen}, {V{\'a}zquez Acosta}, {Vecchi},
  {Velasco}, {Vialle}, {Vitale}, {Vitillo}, {Wang}, {Wang}, {Wang}, {Wang},
  {Wang}, {Wang}, {Wei}, {Weng}, {Whitman}, {Wu}, {Wu}, {Xiong}, {Xu}, {Yan},
  {Yang}, {Yang}, {Yang}, {Yi}, {Yu}, {Yu}, {Zannoni}, {Zeissler}, {Zhang},
  {Zhang}, {Zhang}, {Zhang}, {Zhang}, {Zhang}, {Zheng}, {Zhuang}, {Zhukov},
  {Zichichi}, {Zimmermann}, {Zuccon}, \& {AMS
  Collaboration}}]{2017PhRvL.119y1101A}
{Aguilar}, M., {Ali Cavasonza}, L., {Alpat}, B., {et~al.} 2017, \prl, 119,
  251101

\bibitem[{{Aharonian} {et~al.}(2021){Aharonian}, {An}, {Axikegu}, {Bai}, {Bao},
  {Bastieri}, {Bi}, {Bi}, {Cai}, {Cai}, {Cao}, {Cao}, {Chang}, {Chang},
  {Chang}, {Chen}, {Chen}, {Chen}, {Chen}, {Chen}, {Chen}, {Chen}, {Chen},
  {Chen}, {Chen}, {Chen}, {Chen}, {Chen}, {Cheng}, {Cheng}, {Cui}, {Cui},
  {Cui}, {Dai}, {Dai}, {Dai}, {Danzengluobu}, {Della Volpe}, {D'Ettorre
  Piazzoli}, {Dong}, {Fan}, {Fan}, {Fan}, {Fang}, {Fang}, {Feng}, {Feng},
  {Feng}, {Feng}, {Gao}, {Gao}, {Gao}, {Gao}, {Ge}, {Geng}, {Gong}, {Gou},
  {Gu}, {Guo}, {Guo}, {Guo}, {Guo}, {Han}, {He}, {He}, {He}, {He}, {He}, {He},
  {Heller}, {Hor}, {Hou}, {Hou}, {Hu}, {Hu}, {Hu}, {Hu}, {Huang}, {Huang},
  {Huang}, {Huang}, {Huang}, {Ji}, {Ji}, {Jia}, {Jiang}, {Jiang}, {Jin},
  {Kuleshov}, {Levochkin}, {Li}, {Li}, {Li}, {Li}, {Li}, {Li}, {Li}, {Li},
  {Li}, {Li}, {Li}, {Li}, {Li}, {Li}, {Li}, {Li}, {Li}, {Liang}, {Liang},
  {Lin}, {Liu}, {Liu}, {Liu}, {Liu}, {Liu}, {Liu}, {Liu}, {Liu}, {Liu}, {Liu},
  {Liu}, {Liu}, {Liu}, {Liu}, {Liu}, {Long}, {Lu}, {Lv}, {Ma}, {Ma}, {Ma},
  {Mao}, {Masood}, {Mitthumsiri}, {Montaruli}, {Nan}, {Pang},
  {Pattarakijwanich}, {Pei}, {Qi}, {Ruffolo}, {Rulev}, {S{\'a}iz}, {Shao},
  {Shchegolev}, {Sheng}, {Shi}, {Song}, {Stenkin}, {Stepanov}, {Sun}, {Sun},
  {Sun}, {Tam}, {Tang}, {Tian}, {Wang}, {Wang}, {Wang}, {Wang}, {Wang}, {Wang},
  {Wang}, {Wang}, {Wang}, {Wang}, {Wang}, {Wang}, {Wang}, {Wang}, {Wang},
  {Wang}, {Wang}, {Wang}, {Wang}, {Wang}, {Wang}, {Wei}, {Wei}, {Wei}, {Wen},
  {Wu}, {Wu}, {Wu}, {Wu}, {Wu}, {Xi}, {Xia}, {Xia}, {Xiang}, {Xiao}, {Xiao},
  {Xin}, {Xin}, {Xing}, {Xu}, {Xu}, {Xue}, {Yan}, {Yang}, {Yang}, {Yang},
  {Yang}, {Yang}, {Yang}, {Yang}, {Yao}, {Yao}, {Ye}, {Yin}, {Yin}, {You},
  {You}, {Yu}, {Yuan}, {Zeng}, {Zeng}, {Zeng}, {Zeng}, {Zha}, {Zhai}, {Zhang},
  {Zhang}, {Zhang}, {Zhang}, {Zhang}, {Zhang}, {Zhang}, {Zhang}, {Zhang},
  {Zhang}, {Zhang}, {Zhang}, {Zhang}, {Zhang}, {Zhang}, {Zhang}, {Zhang},
  {Zhang}, {Zhang}, {Zhao}, {Zhao}, {Zhao}, {Zhao}, {Zhao}, {Zheng}, {Zheng},
  {Zhou}, {Zhou}, {Zhou}, {Zhou}, {Zhou}, {Zhou}, {Zhu}, {Zhu}, {Zhu}, {Zhu},
  {Zuo}, {LHAASO Collaboration}, \& {Huang}}]{2021PhRvL.126x1103A}
{Aharonian}, F., {An}, Q., {Axikegu}, Bai, L.~X., {et~al.} 2021, \prl, 126,
  241103

\bibitem[{{Ahlers}(2016)}]{2016PhRvL.117o1103A}
{Ahlers}, M. 2016, Physical Review Letters, 117, 151103

\bibitem[{{Ahlers}(2019)}]{2019ApJ...886L..18A}
---. 2019, \apjl, 886, L18

\bibitem[{{Alemanno} {et~al.}(2021){Alemanno}, {An}, {Azzarello}, {Barbato},
  {Bernardini}, {Bi}, {Cai}, {Catanzani}, {Chang}, {Chen}, {Chen}, {Chen},
  {Cui}, {Cui}, {Cui}, {Dai}, {D'Amone}, {de Benedittis}, {de Mitri}, {de
  Palma}, {Deliyergiyev}, {di Santo}, {Dong}, {Dong}, {Donvito}, {Droz},
  {Duan}, {Duan}, {D'Urso}, {Fan}, {Fan}, {Fang}, {Fang}, {Feng}, {Feng},
  {Fusco}, {Gao}, {Gargano}, {Gong}, {Gong}, {Guo}, {Guo}, {Guo}, {Han}, {Hu},
  {Huang}, {Huang}, {Huang}, {Ionica}, {Jiang}, {Kong}, {Kotenko}, {Kyratzis},
  {Lei}, {Li}, {Li}, {Li}, {Li}, {Liang}, {Liu}, {Liu}, {Liu}, {Liu}, {Liu},
  {Liu}, {Loparco}, {Luo}, {Ma}, {Ma}, {Ma}, {Ma}, {Marsella}, {Mazziotta},
  {Mo}, {Niu}, {Pan}, {Parenti}, {Peng}, {Peng}, {Perrina}, {Qiao}, {Rao},
  {Ruina}, {Salinas}, {Shang}, {Shen}, {Shen}, {Shen}, {Silveri}, {Song},
  {Stolpovskiy}, {Su}, {Su}, {Sun}, {Surdo}, {Teng}, {Tykhonov}, {Wang},
  {Wang}, {Wang}, {Wang}, {Wang}, {Wang}, {Wang}, {Wang}, {Wang}, {Wei}, {Wei},
  {Wei}, {Wen}, {Wu}, {Wu}, {Wu}, {Wu}, {Wu}, {Xia}, {Xu}, {Xu}, {Xu}, {Xu},
  {Xue}, {Yang}, {Yang}, {Yang}, {Yao}, {Yu}, {Yuan}, {Yuan}, {Yue}, {Zang},
  {Zhang}, {Zhang}, {Zhang}, {Zhang}, {Zhang}, {Zhang}, {Zhang}, {Zhang},
  {Zhang}, {Zhang}, {Zhao}, {Zhao}, {Zhao}, {Zhou}, {Zhu}, \& {Dampe
  Collaboration}}]{2021PhRvL.126t1102A}
{Alemanno}, F., {An}, Q., {Azzarello}, P., {et~al.} 2021, \prl, 126, 201102

\bibitem[{{Ambrosio} {et~al.}(2003){Ambrosio}, {Antolini}, {Baldini},
  {Barbarino}, {Barish}, {Battistoni}, {Becherini}, {Bellotti}, {Bemporad},
  {Bernardini}, {Bilokon}, {Bower}, {Brigida}, {Bussino}, {Cafagna},
  {Calicchio}, {Campana}, {Carboni}, {Caruso}, {Cecchini}, {Cei}, {Chiarella},
  {Chiarusi}, {Choudhary}, {Coutu}, {Cozzi}, {de Cataldo}, {Dekhissi}, {de
  Marzo}, {de Mitri}, {Derkaoui}, {de Vincenzi}, {di Credico}, {Erriquez},
  {Favuzzi}, {Forti}, {Fusco}, {Giacomelli}, {Giannini}, {Giglietto},
  {Giorgini}, {Grassi}, {Grillo}, {Gustavino}, {Habig}, {Hanson}, {Heinz},
  {Katsavounidis}, {Katsavounidis}, {Kearns}, {Kim}, {Kyriazopoulou},
  {Lamanna}, {Lane}, {Levin}, {Lipari}, {Longley}, {Longo}, {Loparco},
  {Maaroufi}, {Mancarella}, {Mandrioli}, {Margiotta}, {Marini}, {Martello},
  {Marzari-Chiesa}, {Mazziotta}, {Michael}, {Miller}, {Monacelli}, {Montaruli},
  {Monteno}, {Mufson}, {Musser}, {Nicol{\`o}}, {Nolty}, {Orth}, {Osteria},
  {Palamara}, {Patrizii}, {Pazzi}, {Peck}, {Perrone}, {Petrera}, {Popa},
  {Rain{\`o}}, {Reynoldson}, {Ronga}, {Satriano}, {Scapparone}, {Scholberg},
  {Serra}, {Sioli}, {Sirri}, {Sitta}, {Spinelli}, {Spinetti}, {Spurio},
  {Steinberg}, {Stone}, {Sulak}, {Surdo}, {Tarl{\`e}}, {Togo}, {Vakili},
  {Walter}, \& {Webb}}]{2003PhRvD..67d2002A}
{Ambrosio}, M., {Antolini}, R., {Baldini}, A., {et~al.} 2003, \prd, 67, 042002

\bibitem[{{Amenomori} {et~al.}(2005){Amenomori}, {Ayabe}, {Cui},
  {Danzengluobu}, {Ding}, {Ding}, {Feng}, {Feng}, {Gao}, {Geng}, {Guo}, {He},
  {He}, {Hibino}, {Hotta}, {Hu}, {Hu}, {Huang}, {Huang}, {Jia}, {Kajino},
  {Kasahara}, {Katayose}, {Kato}, {Kawata}, {Labaciren}, {Le}, {Li}, {Lu},
  {Lu}, {Meng}, {Mizutani}, {Mu}, {Munakata}, {Nagai}, {Nanjo}, {Nishizawa},
  {Ohnishi}, {Ohta}, {Onuma}, {Ouchi}, {Ozawa}, {Ren}, {Saito}, {Sakata},
  {Sasaki}, {Shibata}, {Shiomi}, {Shirai}, {Sugimoto}, {Takita}, {Tan},
  {Tateyama}, {Torii}, {Tsuchiya}, {Udo}, {Utsugi}, {Wang}, {Wang}, {Wang},
  {Wang}, {Wu}, {Xue}, {Yamamoto}, {Yan}, {Yang}, {Yasue}, {Ye}, {Yu}, {Yuan},
  {Yuda}, {Zhang}, {Zhang}, {Zhang}, {Zhang}, {Zhang}, {Zhang}, {Zhaxisangzhu},
  {Zhou}, \& {Tibet As{\ensuremath{\gamma}}
  Collaboration}}]{2005ApJ...626L..29A}
{Amenomori}, M., {Ayabe}, S., {Cui}, S.~W., {et~al.} 2005, \apjl, 626, L29

\bibitem[{{Amenomori} {et~al.}(2006){Amenomori}, {Ayabe}, {Bi}, {Chen}, {Cui},
  {Danzengluobu}, {Ding}, {Ding}, {Feng}, {Feng}, {Feng}, {Gao}, {Geng}, {Guo},
  {He}, {He}, {Hibino}, {Hotta}, {Hu}, {Hu}, {Huang}, {Huang}, {Jia}, {Kajino},
  {Kasahara}, {Katayose}, {Kato}, {Kawata}, {Labaciren}, {Le}, {Li}, {Li},
  {Lou}, {Lu}, {Lu}, {Meng}, {Mizutani}, {Mu}, {Munakata}, {Nagai}, {Nanjo},
  {Nishizawa}, {Ohnishi}, {Ohta}, {Onuma}, {Ouchi}, {Ozawa}, {Ren}, {Saito},
  {Saito}, {Sakata}, {Sako}, {Sasaki}, {Shibata}, {Shiomi}, {Shirai},
  {Sugimoto}, {Takita}, {Tan}, {Tateyama}, {Torii}, {Tsuchiya}, {Udo}, {Wang},
  {Wang}, {Wang}, {Wang}, {Wu}, {Xue}, {Yamamoto}, {Yan}, {Yang}, {Yasue},
  {Ye}, {Yu}, {Yuan}, {Yuda}, {Zhang}, {Zhang}, {Zhang}, {Zhang}, {Zhang},
  {Zhang}, {Zhaxisangzhu}, {Zhou}, \& {Tibet AS{\ensuremath{\gamma}}
  Collaboration}}]{2006Sci...314..439A}
{Amenomori}, M., {Ayabe}, S., {Bi}, X.~J., {et~al.} 2006, Science, 314, 439

\bibitem[{{Amenomori} {et~al.}(2010){Amenomori}, {Bi}, {Chen}, {Cui},
  {Danzengluobu}, {Ding}, {Ding}, {Fan}, {Feng}, {Feng}, {Feng}, {Gao}, {Geng},
  {Gou}, {Guo}, {He}, {He}, {Hibino}, {Hotta}, {Hu}, {Hu}, {Huang}, {Huang},
  {Jia}, {Jiang}, {Kajino}, {Kasahara}, {Katayose}, {Kato}, {Kawata},
  {Labaciren}, {Le}, {Li}, {Li}, {Li}, {Liu}, {Lou}, {Lu}, {Meng}, {Mizutani},
  {Mu}, {Munakata}, {Nagai}, {Nanjo}, {Nishizawa}, {Ohnishi}, {Ohta}, {Ozawa},
  {Saito}, {Saito}, {Sakata}, {Sako}, {Shibata}, {Shiomi}, {Shirai},
  {Sugimoto}, {Takita}, {Tan}, {Tateyama}, {Torii}, {Tsuchiya}, {Udo}, {Wang},
  {Wang}, {Wang}, {Wang}, {Wu}, {Xue}, {Yamamoto}, {Yan}, {Yang}, {Yasue},
  {Ye}, {Yu}, {Yuan}, {Yuda}, {Zhang}, {Zhang}, {Zhang}, {Zhang}, {Zhang},
  {Zhang}, {Zhang}, {Zhaxisangzhu}, {Zhou}, \& {Tibet AS{\ensuremath{\gamma}}
  Collaboration}}]{2010ApJ...711..119A}
{Amenomori}, M., {Bi}, X.~J., {Chen}, D., {et~al.} 2010, \apj, 711, 119

\bibitem[{{Amenomori} {et~al.}(2017){Amenomori}, {Bi}, {Chen}, {Chen}, {Chen},
  {Cui}, {Danzengluobu}, {Ding}, {Feng}, {Feng}, {Feng}, {Gou}, {Guo}, {He},
  {He}, {Hibino}, {Hotta}, {Hu}, {Hu}, {Huang}, {Jia}, {Jiang}, {Kajino},
  {Kasahara}, {Katayose}, {Kato}, {Kawata}, {Kozai}, {Labaciren}, {Le}, {Li},
  {Li}, {Li}, {Liu}, {Liu}, {Liu}, {Lu}, {Meng}, {Miyazaki}, {Mizutani},
  {Munakata}, {Nakajima}, {Nakamura}, {Nanjo}, {Nishizawa}, {Niwa}, {Ohnishi},
  {Ohta}, {Ozawa}, {Qian}, {Qu}, {Saito}, {Saito}, {Sakata}, {Sako}, {Shao},
  {Shibata}, {Shiomi}, {Shirai}, {Sugimoto}, {Takita}, {Tan}, {Tateyama},
  {Torii}, {Tsuchiya}, {Udo}, {Wang}, {Wu}, {Xue}, {Yamamoto}, {Yamauchi},
  {Yang}, {Yuan}, {Yuda}, {Zhai}, {Zhang}, {Zhang}, {Zhang}, {Zhang}, {Zhang},
  {Zhang}, {Zhaxisangzhu}, {Zhou}, \& {Tibet AS{\ensuremath{\gamma}}
  Collaboration}}]{2017ApJ...836..153A}
---. 2017, \apj, 836, 153

\bibitem[{{An} {et~al.}(2019){An}, {Asfandiyarov}, {Azzarello}, {Bernardini},
  {Bi}, {Cai}, {Chang}, {Chen}, {Chen}, {Chen}, {Chen}, {Cui}, {Cui}, {Dai},
  {D'Amone}, {De Benedittis}, {De Mitri}, {Di Santo}, {Ding}, {Dong}, {Dong},
  {Dong}, {Donvito}, {Droz}, {Duan}, {Duan}, {D'Urso}, {Fan}, {Fan}, {Fang},
  {Feng}, {Feng}, {Fusco}, {Gallo}, {Gan}, {Gao}, {Gargano}, {Gong}, {Gong},
  {Guo}, {Guo}, {Guo}, {Han}, {Hu}, {Huang}, {Huang}, {Huang}, {Ionica},
  {Jiang}, {Jin}, {Kong}, {Lei}, {Li}, {Li}, {Li}, {Li}, {Li}, {Liang},
  {Liang}, {Liao}, {Liu}, {Liu}, {Liu}, {Liu}, {Liu}, {Liu}, {Loparco}, {Luo},
  {Ma}, {Ma}, {Ma}, {Ma}, {Ma}, {Marsella}, {Mazziotta}, {Mo}, {Niu}, {Pan},
  {Peng}, {Peng}, {Qiao}, {Rao}, {Salinas}, {Shang}, {Shen}, {Shen}, {Shen},
  {Song}, {Su}, {Su}, {Sun}, {Surdo}, {Teng}, {Tykhonov}, {Vitillo}, {Wang},
  {Wang}, {Wang}, {Wang}, {Wang}, {Wang}, {Wang}, {Wang}, {Wang}, {Wang},
  {Wang}, {Wang}, {Wang}, {Wei}, {Wei}, {Wei}, {Wen}, {Wu}, {Wu}, {Wu}, {Wu},
  {Wu}, {Xi}, {Xia}, {Xu}, {Xu}, {Xu}, {Xu}, {Xue}, {Yang}, {Yang}, {Yang},
  {Yang}, {Yao}, {Yu}, {Yuan}, {Yue}, {Zang}, {Zhang}, {Zhang}, {Zhang},
  {Zhang}, {Zhang}, {Zhang}, {Zhang}, {Zhang}, {Zhang}, {Zhang}, {Zhang},
  {Zhang}, {Zhang}, {Zhao}, {Zhao}, {Zhao}, {Zhou}, {Zhou}, {Zhu}, {Zhu}, \&
  {Zimmer}}]{2019SciA....5.3793A}
{An}, Q., {Asfandiyarov}, R., {Azzarello}, P., {et~al.} 2019, Science Advances,
  5, eaax3793

\bibitem[{{Antoni} {et~al.}(2005){Antoni}, {Apel}, {Badea}, {Bekk}, {Bercuci},
  {Bl{\"u}mer}, {Bozdog}, {Brancus}, {Chilingarian}, {Daumiller}, {Doll},
  {Engel}, {Engler}, {Fe{\ss}ler}, {Gils}, {Glasstetter}, {Haungs}, {Heck},
  {H{\"o}randel}, {Kampert}, {Klages}, {Maier}, {Mathes}, {Mayer}, {Milke},
  {M{\"u}ller}, {Obenland}, {Oehlschl{\"a}ger}, {Ostapchenko}, {Petcu},
  {Rebel}, {Risse}, {Risse}, {Roth}, {Schatz}, {Schieler}, {Scholz}, {Thouw},
  {Ulrich}, {van Buren}, {Vardanyan}, {Weindl}, {Wochele}, \&
  {Zabierowski}}]{2005APh....24....1A}
{Antoni}, T., {Apel}, W.~D., {Badea}, A.~F., {et~al.} 2005, Astroparticle
  Physics, 24, 1

\bibitem[{{Apel} {et~al.}(2013){Apel}, {Arteaga-Vel{\'a}zquez}, {Bekk},
  {Bertaina}, {Bl{\"u}mer}, {Bozdog}, {Brancus}, {Cantoni}, {Chiavassa},
  {Cossavella}, {Daumiller}, {de Souza}, {Di Pierro}, {Doll}, {Engel},
  {Engler}, {Finger}, {Fuchs}, {Fuhrmann}, {Gils}, {Glasstetter}, {Grupen},
  {Haungs}, {Heck}, {H{\"o}randel}, {Huber}, {Huege}, {Kampert}, {Kang},
  {Klages}, {Link}, {{\L}uczak}, {Ludwig}, {Mathes}, {Mayer}, {Melissas},
  {Milke}, {Mitrica}, {Morello}, {Oehlschl{\"a}ger}, {Ostapchenko}, {Palmieri},
  {Petcu}, {Pierog}, {Rebel}, {Roth}, {Schieler}, {Schoo}, {Schr{\"o}der},
  {Sima}, {Toma}, {Trinchero}, {Ulrich}, {Weindl}, {Wochele}, {Wommer}, \&
  {Zabierowski}}]{2013APh....47...54A}
{Apel}, W.~D., {Arteaga-Vel{\'a}zquez}, J.~C., {Bekk}, K., {et~al.} 2013,
  Astroparticle Physics, 47, 54

\bibitem[{{Apel} {et~al.}(2019){Apel}, {Arteaga-Vel{\'a}zquez}, {Bekk},
  {Bertaina}, {Bl{\"u}mer}, {Bonino}, {Bozdog}, {Brancus}, {Cantoni},
  {Chiavassa}, {Cossavella}, {Daumiller}, {de Souza}, {Di Pierro}, {Doll},
  {Engel}, {Fuhrmann}, {Gherghel-Lascu}, {Gils}, {Glasstetter}, {Grupen},
  {Haungs}, {Heck}, {H{\"o}randel}, {Huege}, {Kampert}, {Kang}, {Klages},
  {Link}, {{\L}uczak}, {Mathes}, {Mayer}, {Milke}, {Mitrica}, {Morello},
  {Oehlschl{\"a}ger}, {Ostapchenko}, {Pierog}, {Rebel}, {Roth}, {Schieler},
  {Schoo}, {Schr{\"o}der}, {Sima}, {Toma}, {Trinchero}, {Ulrich}, {Weindl},
  {Wochele}, \& {Zabierowski}}]{2019ApJ...870...91A}
---. 2019, \apj, 870, 91

\bibitem[{{Atkin} {et~al.}(2017){Atkin}, {Bulatov}, {Dorokhov}, {Gorbunov},
  {Filippov}, {Grebenyuk}, {Karmanov}, {Kovalev}, {Kudryashov}, {Kurganov},
  {Merkin}, {Panov}, {Podorozhny}, {Polkov}, {Porokhovoy}, {Shumikhin},
  {Sveshnikova}, {Tkachenko}, {Tkachev}, {Turundaevskiy}, {Vasiliev}, \&
  {Voronin}}]{2017JCAP...07..020A}
{Atkin}, E., {Bulatov}, V., {Dorokhov}, V., {et~al.} 2017, \jcap, 2017, 020

\bibitem[{{Bartoli} {et~al.}(2013){Bartoli}, {Bernardini}, {Bi}, {Bolognino},
  {Branchini}, {Budano}, {Calabrese Melcarne}, {Camarri}, {Cao}, {Cardarelli},
  {Catalanotti}, {Chen}, {Chen}, {Creti}, {Cui}, {Dai}, {D'Amone},
  {Danzengluobu}, {De Mitri}, {D'Ettorre Piazzoli}, {Di Girolamo}, {Di
  Sciascio}, {Feng}, {Feng}, {Feng}, {Gou}, {Guo}, {He}, {Hu}, {Hu},
  {Iacovacci}, {Iuppa}, {Jia}, {Labaciren}, {Li}, {Liguori}, {Liu}, {Liu},
  {Liu}, {Lu}, {Ma}, {Mancarella}, {Mari}, {Marsella}, {Martello},
  {Mastroianni}, {Montini}, {Ning}, {Panareo}, {Panico}, {Perrone}, {Pistilli},
  {Ruggieri}, {Salvini}, {Santonico}, {Sbano}, {Shen}, {Sheng}, {Shi}, {Surdo},
  {Tan}, {Vallania}, {Vernetto}, {Vigorito}, {Wang}, {Wu}, {Wu}, {Xue}, {Yan},
  {Yang}, {Yang}, {Yao}, {Yuan}, {Zha}, {Zhang}, {Zhang}, {Zhang}, {Zhang},
  {Zhaxiciren}, {Zhaxisangzhu}, {Zhou}, {Zhu}, {Zhu}, \&
  {Zizzi}}]{2013PhRvD..88h2001B}
{Bartoli}, B., {Bernardini}, P., {Bi}, X.~J., {et~al.} 2013, prd, 88, 082001

\bibitem[{{Bartoli} {et~al.}(2015){Bartoli}, {Bernardini}, {Bi}, {Cao},
  {Catalanotti}, {Chen}, {Chen}, {Cui}, {Dai}, {D'Amone}, {Danzengluobu}, {De
  Mitri}, {D'Ettorre Piazzoli}, {Di Girolamo}, {Di Sciascio}, {Feng}, {Feng},
  {Feng}, {Gao}, {Gou}, {Guo}, {He}, {Hu}, {Hu}, {Iacovacci}, {Iuppa}, {Jia},
  {Labaciren}, {Li}, {Liu}, {Liu}, {Liu}, {Lu}, {Ma}, {Ma}, {Mancarella},
  {Mari}, {Marsella}, {Mastroianni}, {Montini}, {Ning}, {Perrone}, {Pistilli},
  {Salvini}, {Santonico}, {Shen}, {Sheng}, {Shi}, {Surdo}, {Tan}, {Vallania},
  {Vernetto}, {Vigorito}, {Wang}, {Wu}, {Wu}, {Xue}, {Yang}, {Yang}, {Yao},
  {Yuan}, {Zha}, {Zhang}, {Zhang}, {Zhang}, {Zhang}, {Zhao}, {Zhaxiciren},
  {Zhaxisangzhu}, {Zhou}, {Zhu}, {Zhu}, \& {ARGO-YBJ
  Collaboration}}]{2015ApJ...809...90B}
---. 2015, apj, 809, 90

\bibitem[{{Bartoli} {et~al.}(2018){Bartoli}, {Bernardini}, {Bi}, {Cao},
  {Catalanotti}, {Chen}, {Chen}, {Cui}, {Dai}, {D'Amone}, {Danzengluobu}, {De
  Mitri}, {D'Ettorre Piazzoli}, {Di Girolamo}, {Di Sciascio}, {Feng}, {Feng},
  {Gao}, {Gou}, {Guo}, {He}, {Hu}, {Hu}, {Iacovacci}, {Iuppa}, {Jia},
  {Labaciren}, {Li}, {Liu}, {Liu}, {Liu}, {Lu}, {Ma}, {Ma}, {Mancarella},
  {Mari}, {Marsella}, {Mastroianni}, {Montini}, {Ning}, {Perrone}, {Pistilli},
  {Ruffolo}, {Salvini}, {Santonico}, {Shen}, {Sheng}, {Shi}, {Surdo}, {Tan},
  {Vallania}, {Vernetto}, {Vigorito}, {Wang}, {Wu}, {Wu}, {Xue}, {Yang},
  {Yang}, {Yao}, {Yuan}, {Zha}, {Zhang}, {Zhang}, {Zhang}, {Zhang}, {Zhao},
  {Zhaxiciren}, {Zhaxisangzhu}, {Zhou}, {Zhu}, {Zhu}, \& {ARGO-YBJ
  Collaboration}}]{2018ApJ...861...93B}
---. 2018, \apj, 861, 93

\bibitem[{{Bernard} {et~al.}(2012){Bernard}, {Delahaye}, {Salati}, \&
  {Taillet}}]{2012A&A...544A..92B}
{Bernard}, G., {Delahaye}, T., {Salati}, P., \& {Taillet}, R. 2012, Astronomy
  and Astrophysics, 544, A92

\bibitem[{{Blasi} \& {Amato}(2012)}]{2012JCAP...01..011B}
{Blasi}, P., \& {Amato}, E. 2012, jcap, 1, 11

\bibitem[{{Case} \& {Bhattacharya}(1996)}]{1996A&AS..120C.437C}
{Case}, G., \& {Bhattacharya}, D. 1996, \aaps, 120, 437

\bibitem[{{Cerri} {et~al.}(2017){Cerri}, {Gaggero}, {Vittino}, {Evoli}, \&
  {Grasso}}]{2017JCAP...10..019C}
{Cerri}, S.~S., {Gaggero}, D., {Vittino}, A., {Evoli}, C., \& {Grasso}, D.
  2017, jcap, 10, 019

\bibitem[{{Chiavassa} {et~al.}(2015){Chiavassa}, {Apel},
  {Arteaga-Vel{\'a}zquez}, {Bekk}, {Bertaina}, {Bl{\"u}mer}, {Bozdog},
  {Brancus}, {Cantoni}, {Cossavella}, {Daumiller}, {de Souza}, {di Pierro},
  {Doll}, {Engel}, {Fuhrmann}, {Gherghel-Lascu}, {Gils}, {Glasstetter},
  {Grupen}, {Haungs}, {Heck}, {H{\"o}randel}, {Huber}, {Huege}, {Kampert},
  {Kang}, {Klages}, {Link}, {{\L}uczak}, {Mathes}, {Mayer}, {Milke}, {Mitrica},
  {Morello}, {Oehlschl{\"a}ger}, {Ostapchenko}, {Palmieri}, {Pierog}, {Rebel},
  {Roth}, {Schieler}, {Schoo}, {Schr{\"o}der}, {Sima}, {Toma}, {Trinchero},
  {Ulrich}, {Weindl}, {Wochele}, {Zabierowski}, \& {KASCADE-Grande
  Collaboration}}]{2015ICRC...34..281C}
{Chiavassa}, A., {Apel}, W.~D., {Arteaga-Vel{\'a}zquez}, J.~C., {et~al.} 2015,
  in International Cosmic Ray Conference, Vol.~34, 34th International Cosmic
  Ray Conference (ICRC2015), 281

\bibitem[{Evoli {et~al.}(2008)Evoli, Gaggero, Grasso, \&
  Maccione}]{1475-7516-2008-10-018}
Evoli, C., Gaggero, D., Grasso, D., \& Maccione, L. 2008, Journal of Cosmology
  and Astroparticle Physics, 2008, 018

\bibitem[{{Faherty} {et~al.}(2007){Faherty}, {Walter}, \&
  {Anderson}}]{2007Ap&SS.308..225F}
{Faherty}, J., {Walter}, F.~M., \& {Anderson}, J. 2007, \apss, 308, 225

\bibitem[{{Feng} {et~al.}(2016){Feng}, {Tomassetti}, \&
  {Oliva}}]{2016PhRvD..94l3007F}
{Feng}, J., {Tomassetti}, N., \& {Oliva}, A. 2016, \prd, 94, 123007

\bibitem[{{Fornieri} {et~al.}(2021){Fornieri}, {Gaggero}, {Guberman},
  {Brahimi}, {Luque}, \& {Marcowith}}]{2021PhRvD.104j3013F}
{Fornieri}, O., {Gaggero}, D., {Guberman}, D., {et~al.} 2021, \prd, 104, 103013

\bibitem[{{Frisch} {et~al.}(2015){Frisch}, {Berdyugin}, {Piirola}, {Magalhaes},
  {Seriacopi}, {Wiktorowicz}, {Andersson}, {Funsten}, {McComas}, {Schwadron},
  {Slavin}, {Hanson}, \& {Fu}}]{2015ApJ...814..112F}
{Frisch}, P.~C., {Berdyugin}, A., {Piirola}, V., {et~al.} 2015, \apj, 814, 112

\bibitem[{{Funsten} {et~al.}(2013){Funsten}, {DeMajistre}, {Frisch},
  {Heerikhuisen}, {Higdon}, {Janzen}, {Larsen}, {Livadiotis}, {McComas},
  {M{\"o}bius}, {Reese}, {Reisenfeld}, {Schwadron}, \&
  {Zirnstein}}]{2013ApJ...776...30F}
{Funsten}, H.~O., {DeMajistre}, R., {Frisch}, P.~C., {et~al.} 2013, \apj, 776,
  30

\bibitem[{{Giacalone} \& {Jokipii}(1999)}]{1999ApJ...520..204G}
{Giacalone}, J., \& {Jokipii}, J.~R. 1999, \apj, 520, 204

\bibitem[{{Giacinti} \& {Kirk}(2017)}]{2017ApJ...835..258G}
{Giacinti}, G., \& {Kirk}, J.~G. 2017, \apj, 835, 258

\bibitem[{{Guillian} {et~al.}(2007){Guillian}, {Hosaka}, {Ishihara}, {Kameda},
  {Koshio}, {Minamino}, {Mitsuda}, {Miura}, {Moriyama}, {Nakahata}, {Namba},
  {Obayashi}, {Ogawa}, {Shiozawa}, {Suzuki}, {Takeda}, {Takeuchi}, {Yamada},
  {Higuchi}, {Ishitsuka}, {Kajita}, {Kaneyuki}, {Mitsuka}, {Nakayama},
  {Nishino}, {Okada}, {Okumura}, {Saji}, {Takenaga}, {Desai}, {Kearns},
  {Stone}, {Sulak}, {Wang}, {Goldhaber}, {Casper}, {Gajewski}, {Griskevich},
  {Kropp}, {Liu}, {Mine}, {Smy}, {Sobel}, {Vagins}, {Ganezer}, {Hill}, {Keig},
  {Scholberg}, {Walter}, {Ellsworth}, {Tasaka}, {Kibayashi}, {Learned},
  {Matsuno}, {Messier}, {Hayato}, {Ichikawa}, {Ishida}, {Ishii}, {Iwashita},
  {Kobayashi}, {Nakadaira}, {Nakamura}, {Nitta}, {Oyama}, {Totsuka}, {Suzuki},
  {Hasegawa}, {Kato}, {Maesaka}, {Nakaya}, {Nishikawa}, {Sato}, {Yamamoto},
  {Yokoyama}, {Haines}, {Dazeley}, {Hatakeyama}, {Svoboda}, {Blaufuss},
  {Goodman}, {Sullivan}, {Turcan}, {Habig}, {Fukuda}, {Itow}, {Sakuda},
  {Yoshida}, {Kim}, {Yoo}, {Okazawa}, {Ishizuka}, {Jung}, {Kato}, {Kobayashi},
  {Malek}, {Mauger}, {McGrew}, {Sharkey}, {Yanagisawa}, {Gando}, {Hasegawa},
  {Inoue}, {Shirai}, {Suzuki}, {Nishijima}, {Ishino}, {Watanabe}, {Koshiba},
  {Kielczewska}, {Berns}, {Gran}, {Shiraishi}, {Stachyra}, {Washburn},
  {Wilkes}, \& {Munakata}}]{2007PhRvD..75f2003G}
{Guillian}, G., {Hosaka}, J., {Ishihara}, K., {et~al.} 2007, \prd, 75, 062003

\bibitem[{{Guo} {et~al.}(2016){Guo}, {Tian}, \& {Jin}}]{2016ApJ...819...54G}
{Guo}, Y.-Q., {Tian}, Z., \& {Jin}, C. 2016, \apj, 819, 54

\bibitem[{{Guo} \& {Yuan}(2018)}]{2018PhRvD..97f3008G}
{Guo}, Y.-Q., \& {Yuan}, Q. 2018, \prd, 97, 063008

\bibitem[{{H{\"o}randel}(2003)}]{2003APh....19..193H}
{H{\"o}randel}, J.~R. 2003, Astroparticle Physics, 19, 193

\bibitem[{{Kuhlen} {et~al.}(2022){Kuhlen}, {Phan}, \&
  {Mertsch}}]{2022ApJ...927..110K}
{Kuhlen}, M., {Phan}, V. H.~M., \& {Mertsch}, P. 2022, \apj, 927, 110

\bibitem[{{Liu} {et~al.}(2017){Liu}, {Bi}, {Lin}, {Wang}, \&
  {Yin}}]{2017PhRvD..96b3006L}
{Liu}, W., {Bi}, X.-J., {Lin}, S.-J., {Wang}, B.-B., \& {Yin}, P.-F. 2017, prd,
  96, 023006

\bibitem[{{Liu} {et~al.}(2019){Liu}, {Guo}, \& {Yuan}}]{2019JCAP...10..010L}
{Liu}, W., {Guo}, Y.-Q., \& {Yuan}, Q. 2019, jcap, 2019, 010

\bibitem[{{Liu} {et~al.}(2020){Liu}, {Lin}, {Hu}, {Guo}, \&
  {Li}}]{2020ApJ...892....6L}
{Liu}, W., {Lin}, S.-j., {Hu}, H.-b., {Guo}, Y.-q., \& {Li}, A.-f. 2020, \apj,
  892, 6

\bibitem[{{Liu} {et~al.}(2018){Liu}, {Yao}, \& {Guo}}]{2018ApJ...869..176L}
{Liu}, W., {Yao}, Y.-h., \& {Guo}, Y.-Q. 2018, \apj, 869, 176

\bibitem[{{Manchester} {et~al.}(2005){Manchester}, {Hobbs}, {Teoh}, \&
  {Hobbs}}]{2005AJ....129.1993M}
{Manchester}, R.~N., {Hobbs}, G.~B., {Teoh}, A., \& {Hobbs}, M. 2005, \aj, 129,
  1993

\bibitem[{{McComas} {et~al.}(2009){McComas}, {Allegrini}, {Bochsler},
  {Bzowski}, {Christian}, {Crew}, {DeMajistre}, {Fahr}, {Fichtner}, {Frisch},
  {Funsten}, {Fuselier}, {Gloeckler}, {Gruntman}, {Heerikhuisen}, {Izmodenov},
  {Janzen}, {Knappenberger}, {Krimigis}, {Kucharek}, {Lee}, {Livadiotis},
  {Livi}, {MacDowall}, {Mitchell}, {M{\"o}bius}, {Moore}, {Pogorelov},
  {Reisenfeld}, {Roelof}, {Saul}, {Schwadron}, {Valek}, {Vanderspek}, {Wurz},
  \& {Zank}}]{2009Sci...326..959M}
{McComas}, D.~J., {Allegrini}, F., {Bochsler}, P., {et~al.} 2009, Science, 326,
  959

\bibitem[{{Mertsch}(2011)}]{2011JCAP...02..031M}
{Mertsch}, P. 2011, Journal of Cosmology and Astroparticle Physics, 2011, 031

\bibitem[{{Qiao} {et~al.}(2019){Qiao}, {Liu}, {Guo}, \&
  {Yuan}}]{2019JCAP...12..007Q}
{Qiao}, B.-Q., {Liu}, W., {Guo}, Y.-Q., \& {Yuan}, Q. 2019, jcap, 2019, 007

\bibitem[{{Schwadron} {et~al.}(2014){Schwadron}, {Adams}, {Christian},
  {Desiati}, {Frisch}, {Funsten}, {Jokipii}, {McComas}, {Moebius}, \&
  {Zank}}]{2014Sci...343..988S}
{Schwadron}, N.~A., {Adams}, F.~C., {Christian}, E.~R., {et~al.} 2014, Science,
  343, 988

\bibitem[{{Tian} {et~al.}(2020){Tian}, {Liu}, {Yang}, {Fu}, {Xu}, {Yao}, \&
  {Guo}}]{2020ChPhC..44h5102T}
{Tian}, Z., {Liu}, W., {Yang}, B., {et~al.} 2020, Chinese Physics C, 44, 085102

\bibitem[{{Tomassetti}(2012)}]{2012ApJ...752L..13T}
{Tomassetti}, N. 2012, apj, 752, L13

\bibitem[{{Tomassetti}(2015)}]{2015PhRvD..92h1301T}
---. 2015, \prd, 92, 081301

\bibitem[{{Yoon} {et~al.}(2017){Yoon}, {Anderson}, {Barrau}, {Conklin},
  {Coutu}, {Derome}, {Han}, {Jeon}, {Kim}, {Kim}, {Lee}, {Lee}, {Lee}, {Lee},
  {Link}, {Menchaca-Rocha}, {Mitchell}, {Mognet}, {Nutter}, {Park},
  {Picot-Clemente}, {Putze}, {Seo}, {Smith}, \& {Wu}}]{2017ApJ...839....5Y}
{Yoon}, Y.~S., {Anderson}, T., {Barrau}, A., {et~al.} 2017, \apj, 839, 5

\bibitem[{{Yuan} {et~al.}(2020){Yuan}, {Qiao}, {Guo}, {Fan}, \&
  {Bi}}]{2020FrPhy..1624501Y}
{Yuan}, Q., {Qiao}, B.-Q., {Guo}, Y.-Q., {Fan}, Y.-Z., \& {Bi}, X.-J. 2020,
  Frontiers of Physics, 16, 24501

\bibitem[{{Zhang} {et~al.}(2022){Zhang}, {Liu}, \&
  {Zeng}}]{2022MNRAS.511.6218Z}
{Zhang}, Y., {Liu}, S., \& {Zeng}, H. 2022, \mnras, 511, 6218

\bibitem[{{Zhao} {et~al.}(2022){Zhao}, {Liu}, {Yuan}, {Hu}, {Bi}, {Wu}, {Zhou},
  \& {Guo}}]{2022ApJ...926...41Z}
{Zhao}, B., {Liu}, W., {Yuan}, Q., {et~al.} 2022, \apj, 926, 41

\end{thebibliography}
%%%%%%%%%%%%%%%%%%%%%%%%%%%%%%%%%%%%%%%%%%%%%%%%%%%%%%%%%%%%%%%%%%%%%%

\end{document}